\documentclass[journal]{IEEEtran}
\usepackage{cite}
\ifCLASSINFOpdf
  \usepackage[pdftex]{graphicx}
\else
\fi
\usepackage{amsmath}
\usepackage{algorithmic}
\usepackage{array}
\ifCLASSOPTIONcompsoc
  \usepackage[caption=false,font=normalsize,labelfont=sf,textfont=sf]{subfig}
\else
  \usepackage[caption=false,font=footnotesize]{subfig}
\fi
\usepackage{fixltx2e}
\usepackage{stfloats}
\usepackage{url}
\begin{document}
\title{Simulation Comparisons of Vehicle-based and Phase-based Traffic Control for Autonomous Vehicles at Isolated Intersections}

\author{Chen~Yang,
        Xi~Lin,
        Meng~Li,
        and~Fang~He % <-this % stops a space
\thanks{Manuscript received August 12, 2020; revised Month dd, yyyy. \textit{(Corresponding author: Meng Li.)}}
\thanks{C. Yang is with the Department of Civil Engineering, Tsinghua University, Beijing 100084, P.R. China, e-mail: (yang-c19@mails.tsinghua.edu.cn).}% <-this % stops a space
\thanks{X. Lin is with the Department of Civil Engineering, Tsinghua University, Beijing 100084, P.R. China, e-mail: (x-lin@mail.tsinghua.edu.cn).}% <-this % stops a space
\thanks{M. Li is with the Department of Civil Engineering, Tsinghua University, Beijing 100084, P.R. China, e-mail: (mengli@tsinghua.edu.cn).}% <-this % stops a space
\thanks{F. He is with the Department of Industrial Engineering, Tsinghua University, Beijing 100084, P.R. China, e-mail: (fanghe@tsinghua.edu.cn).}% <-this % stops a space
}


\ifCLASSOPTIONpeerreview
\else
 \markboth{IEEE TRANSACTIONS ON INTELLIGENT TRANSPORTATION SYSTEMS,~Vol.~xx, No.~xx, Month~2020} %
 {Yang \MakeLowercase{\textit{et al.}}: Simulation Comparisons of Vehicle-based and Phase-based Traffic Control for Autonomous Vehicles at Isolated Intersections}
\fi



\maketitle

\begin{abstract}
With the advent of autonomous driving technologies, traffic control at intersections is expected to experience revolutionary changes. Various novel intersection control methods have been proposed in the existing literature, and they can be roughly divided into two categories: vehicle-based traffic control and phase-based traffic control. Phase-based traffic control can be treated as updated versions of the current intersection signal control with the incorporation of the performance of autonomous vehicle functions. Meanwhile, vehicle-based traffic control utilizes some brand-new methods, mostly in real-time fashion, to organize traffic at intersections for safe and efficient vehicle passages. However, to date, no systematic comparison between these two control categories has been performed to suggest their advantages and disadvantages. This paper conducts a series of numerical simulations under various traffic scenarios to perform a fair comparison of their performances. Specifically, we allow trajectory adjustments of incoming vehicles under phase-based traffic control, while for its vehicle-based counterpart, we implement two strategies, i.e., the first-come-first-serve strategy and the conflict-point based rolling-horizon optimization strategy. Overall, the simulation results show that vehicle-based traffic control generally incurs a negligible delay when traffic demand is low but lead to an excessive queuing time as the traffic volume becomes high. However, performance of vehicle-based traffic control may benefit from reduction in conflicting vehicle pairs. We also discovered that when autonomous driving technologies are not mature, the advantages of phase-based traffic control are much more distinct.

\end{abstract}


% \fi
\IEEEpeerreviewmaketitle

\section{Introduction}
\IEEEPARstart{U}{rban} road systems contain a large number of intersections. At intersections, the trajectories of vehicles traveling from multiple directions conflict with each other, posing a risk of collisions. The introduction of traffic signals has contributed significantly to intersection collision avoidance, but it has also made intersections the bottleneck of the road traffic network. Delays will inevitably occur when vehicles encounter red lights, and the startup loss as well as the phase transition loss also lower the intersection capacity. Many previous studies have focused on minimizing delay by adapting signal timing according to the estimated traffic demand~\cite{Mirchandani2001,Lin2004}, but the implementation of these schemes is difficult and the potential improvement provided by these methods may be limited in a human-driven environment. \par

With the rapid development of artificial intelligence and wireless communications, connected automated vehicle (CAV) technology is considered to be one of the most promising fields in future transportation. CAVs are able to interact with other vehicles on the road as well as roadside facilities, leading to improved driving trajectories to minimize travel delay, fuel consumption and network throughput~\cite{Chen2016,Tajalli2018}. Moreover, benefiting from sensors installed onboard and inter-vehicle communication, vehicles can measure the headway in a more accurate and timely manner and thus make decisions much more effectively, allowing CAVs to maintain a shorter headway~\cite{Arem2006,Wang2018}. Due to all of the desirable characteristics of CAVs, the form of traffic organization, particularly at intersections, may experience revolutionary changes in the coming years. Studies performed to date have explored a variety of isolated intersection control methodologies, which can be roughly categorized into vehicle-based and phase-based schemes. Vehicle based control determines the passing orders and trajectories for specific vehicles, while phase-based strategies are adaptive for typical traffic demands. Below, we provide a brief review of these control methods. \par

\subsection{Vehicle-based traffic control} \label{intro_SigFree_subsec}

Leveraging the connectivity and controllability of CAVs, intersection traffic signals can be totally abandoned, and centralized or decentralized controllers can be placed on intersections to detect the approaching CAV and arrange movements for it. Following this approach, various control concepts have been proposed. According to the study of Meng \textit{et al.}~\cite{Meng2018}, the signal-free vehicle-based intersection control can be categorized into two kinds: "ad-hoc negotiation based"~\cite{Dresner2004,Dresner2008,Li2013} and "planning based"~\cite{Zhu2015}. With the ad-hoc negotiation based methods, intersections are mainly organized under the rule of "first come first served"(FCFS)~\cite{Meng2018}, while planning based methods usually utilize optimization or searching approaches to determine the passing trajectories of multiple vehicles.\par

One of the earliest concepts of ad-hoc negotiation based intersection controls is to allow CAVs to make passing reservations while approaching an intersection, with the centralized intersection manager ensuring conflict-free outcomes. In this approach, Dresner and Stone~\cite{Dresner2004} proposed a multiagent FCFS intersection control policy, and it has been widely studied in subsequent researches~\cite{Dresner2008,Fajardo2011}. The control divides the intersection into multiple tiles, and by applying the restriction that a given tile can only be occupied by one CAV at any given time, collision avoidance is realized. These studies claim to achieve a lower delay for the FCFS policy in a single intersection, and the simulation conducted in VISSIM by Li \textit{et al.}~\cite{Li2013} has also verified that the FCFS policy outperforms traditional signal control. According to the experiments in~\cite{Carlino2013}, the intersection efficiency could be further slightly improved by introducing an auction-based system. \par

In a different approach, planning based intersection control can handle multiple vehicles at the same time instead of arranging vehicle trajectories exactly in the arrival order. In this case, the controller would perceive vehicles that arrive within a certain time period and determine their passing orders and trajectories by solving an optimization problem. Some existing studies achieve collision avoidance by preventing vehicles with intersecting trajectories from being in the intersection concurrently~\cite{Zhu2015,Meng2018}. In~\cite{Zhu2015}, a discrete time model (linear programming formulation for autonomous intersection control, or LPAIC) is formulated to achieve minimum traffic delay in a 4-leg, 4-lane intersection, with the outputs from each direction satisfying the demands. Meng \textit{et al.}~\cite{Meng2018} studied a simpler intersection scenario with only one lane in each leg. Both planning based and ad-hoc negotiation based frameworks are adopted to organize the vehicle passing order, and the simulations show that the planning-based approach is superior to the ad-hoc negotiation based approach with respect to both average value and standard deviation of vehicle delay, particularly when the traffic demand is high. In a more elaborate method, Lee and Park~\cite{Lee2012} proposed a cooperative vehicle intersection control (CVIC) system to adjust the acceleration behavior of passing vehicles by solving nonlinear constrained programming problems to minimize the overlapping length of the intersected trajectories. Moreover, to achieve a higher traffic efficiency, some studies allow vehicles with intersecting trajectories to enter the intersection simultaneously, as long as they do not pass the intersected point at the same time~\cite{Levin2017Confilct,Levin2017On}. In the conflict point intersection control (CPIC) model~\cite{Levin2017Confilct}, the spatial trajectories of vehicles passing the intersection are predesigned, and the conflict points are accordingly defined as the nodes where two trajectories intercept. The model then introduces a mixed integer program to optimize the entering time and passing speed of each vehicle. \par

\subsection{Phase-based traffic control} \label{intro_Sig_subsec}

Phase-based traffic control alternates passing permissions among conflicting traffic movements to provide a safe as well as efficient intersection organization. In a control cycle, conflict movements are instructed to pass the intersection in different phases, and the instructions are delivered to drivers using a set of signal lights. The assignment of phase length, in general, focuses on the overall traffic demand and the demand distribution pattern among conflict movements. One representative phase-based traffic control which has been implemented in current intersections is the pretimed signalized control. Unlike the actuated or semi-actuated signalized control which uses detectors to adjust phase settings based on real-time detected vehicles and pedestrians, in pretimed signalized control, the phase length is fixed based on typical traffic characteristics (e.g., traffic demands or the headway in average). In certain cases, a fluctuation of arrival rate on some directions might cause temporary queuing, but it should finally dissipate as long as the traffic flow characteristics remain stable. \par

In the era of autonomous driving, the phase-based intersection control may develop in other forms with higher traffic efficiency. While in this approach we still borrow the terminology of traditional signal control, e.g., pretimed phases and green/red lights, but it should be noted that the actual traffic lights are not required; rather, phase-based traffic control under the CAV environment is essentially a “collective” approach of traffic organization that assigns passing allowances to a group of non-conflicting vehicle movements in the same period of time. From the vehicle’s perspective, dynamic speed advice ensuring that vehicles pass through the signalized intersection at the maximum allowable speed during the green duration has been verified to be able to considerably reduce fuel~\cite{Trayford1984,Sanchez2006} or electricity~\cite{Wu2015} consumption. Moreover, in a pure CAV environment, the centralized traffic manager can ensure the nonstop passing of all vehicles, reducing the total phase transition losses of the intersection; as a result, the cycle length can be strongly decreased.~\cite{Zhou2017} and~\cite{Ma2017} proposed a parsimonious shooting heuristic to optimize the detailed trajectories of multiple CAVs approaching an intersection simultaneously, and following this research direction, Li \textit{et al.}~\cite{Li2018} simplified the trajectory optimization approach and lowered the computational complexity while preserving most of the desirable features of the former model. These studies reveal that the pretimed phase-based strategy can also be a practical intersection control method in the coming autonomous driving era; compared to its vehicle-based counterpart, phase-based traffic control is easier to implement, and the computational burden added to the controller is strongly alleviated since it is not necessary to solve complicated mathematical programming problems. \par

\vspace{3 ex}

Only limited work comparing the performance characteristics of the above two control philosophies (i.e., the vehicle-based control and the phase-based control) has been reported. Most previous studies have failed to consider the use of CAV technologies to improve the performance of the signalized control \cite{Dresner2008,Li2013}, leading to an underestimation of its potential to some extent. Recently, some researchers have observed that the vehicle-based control cannot always outperform the conventional signal control. Levin \textit{et al.}\cite{Levin2016} indicated some traffic scenarios for which signalized intersection controls outperform vehicle-based controls, and Patel \textit{et al.}\cite{Patel2019} investigated the optimal placement of vehicle-based and signalized intersections in urban networks. Nevertheless, the existing comparisons have been performed mostly under limited traffic scenarios, which are incapable of producing convincing and comprehensive conclusions. \par

Intuitively, compared to phase-based controls such as the signalized control, vehicle-based controls induce more “crossing-type interactions”, i.e., two vehicles on two conflicting lanes pass through the conflict point consecutively, and the crossing-type interaction generally requires a large time gap to ensure safety \cite{Yu2019}. Based on this intuition, the use of vehicle-based controls may lead to lower vehicle delays for light traffic, but one can naturally question their capabilities in a high traffic environment. For a fair comparison of different intersection control schemes in the era of CAVs, this study conducts numerical analyses on three intersection control protocols, i.e., pretimed phase-based control (PPC), vehicle-based control with the FCFS strategy and vehicle-based control with the CPIC strategy \cite{Levin2017Confilct}, under heterogeneous traffic demand patterns and intersection layouts. Specifically, in the PPC strategy, the approaching vehicles’ trajectories are adjusted to coordinate the green phases; the CPIC strategy requires the solution of mixed-integer programming models in real-time fashion, and to simulate its practical usage, we only allow a limited computational time. The testing scenarios include a variety of demand patterns from light to heavy, from balanced to imbalanced (in terms of arrival rates from different legs), and from stable to fluctuating. The tests are conducted with different intersection layouts, including four-leg and three-leg intersections. Additionally, we also incorporate scenarios with different levels of technological maturity to validate the performance characteristics of the two control philosophies. \par

The remainder of this paper is organized as follows. Section \ref{method_sec} describes the control models compared in this paper. In Section \ref{simulation_sec}, we simulate the control models under various traffic demand scenarios and intersection layouts and describe the simulation results. Finally, we conclude by presenting our findings in Section \ref{concluding_sec}.\par

\section{Control strategies} \label{method_sec}

\begin{figure}[!t]
    \centerline{\includegraphics[width=0.45\textwidth]{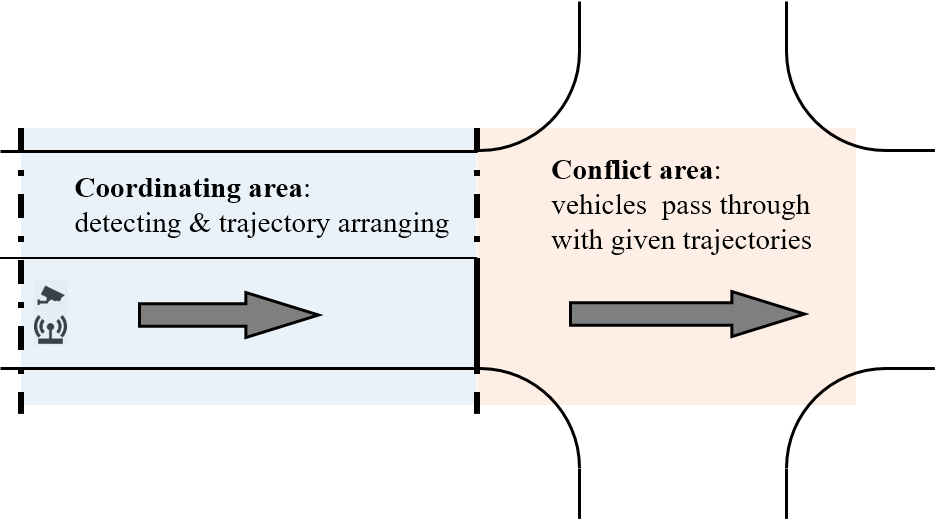}}
    \caption{Layout of the control segment\label{Fig1}}
\end{figure}

This study focuses on control strategy comparisons for a isolated intersection under the CAV environment. As shown in Fig. \ref{Fig1}, the investigated intersection area includes the intersection core (the conflict area) and the coordinating area with a length of several hundred meters for each branch. Within the segment, the managing center can obtain the current state and traffic intention of all vehicles through roadside sensors or V2I communication, and then organizes the movements of vehicles. The center will calculate a desired trajectory for each vehicle using a specific control model and send the trajectory to the CAV. We assume that all involved vehicles are capable of understanding and following trajectories within a limited error. Under this setting, we present a thorough comparison among three control models for isolated intersections. The three models are a pretimed phase-based traffic control model and two vehicle-based models, namely, the first-come-first-serve (FCFS) control~\cite{Dresner2008} and the conflict-point intersection control (CPIC)~\cite{Levin2017Confilct}. FCFS is relatively simple and intuitive, and the latter control is expected to obtain better solutions with a much higher computational burden (since it requires the solution of a mixed-integer program). In this section, a brief description of the comparison framework and the three intersection control models is provided.  \par

%Example for bibliography citations cite~\citep{Elbaum2002}, cites~\cite{Allen2011,Yoo2007}

\subsection{Comparison frameworks} \label{method_Lot_subsec}

To fairly compare the three intersection control models, we should guarantee the equality of the parameters involved in the models. Vehicles of the three models share the same performance characteristics, for example, the maximum acceleration/deceleration rate and the reaction time. Also, vehicles would enter the area at the same cruising velocity, and the speed limits are equally set. In addition, the intersection layouts, including the length of the coordinating area and the size of the conflict area, is fairly set. \par

We also guarantee a similar level of safety. For the compared models, we have a same setting of the safe gap value in the conflict area; specially, we adopt different methods to achieve the safe distance. In the phase-based model, we limit the minimum spatial headway between lag and preceding vehicles; the time gap between phases is also considered to avoid collisions among vehicles from different directions. In vehicle-based models, we expand the longitudinal size of vehicles through spatial buffers, and by that means we preserve the safe gap between vehicles. In the coordinating area, vehicles should also maintain a minimum safe gap. We introduce the Gipps' safe distance rule~\cite{Gipps1981} to avoid rear-end collisions. For simplification, we assume that all vehicles enter the coordinating area on the lane corresponding to the desired travel direction, so that no lane changes will be conducted.  \par

The detailed introduction of these control models is presented in following subsections. \par

\subsection{Pretimed phase-based control} \label{method_Sig_subsec}

The first discussed control is the pretimed phase-based control. Under the CAV environment, we can rely on the virtual signal lights to specify the vehicle that receives permission and passes the intersection. Similar to the conventional signalized intersection control, the compared control has pretimed phase settings (4 phases for 4-leg intersections and 3 phases for 3-leg intersections), and a vehicle is allowed to pass through the intersection only when its "signal" is green. The green time in each phase is allocated among the phase cycle and is adapted to the traffic demand and average headway. The optimal cycle length is determined through a grid search by conducting a series of simulations. For each arrival pattern, we generated 10 arrival sequences and simulated them in different cycle lengths: from 16 s to 120 s in 4-leg intersections, and 12 s to 90 s in T-type junctions. The cycle length with the lowest average delay is then selected as the optimal cycle length in this traffic pattern. \par

Under the CAV environment, the phase settings can be sent to CAVs in advance, allowing them to adjust their speed and match their entrance to the intersection with green lights. A desirable trajectory provides smooth acceleration and deceleration for the vehicle, lowering fuel consumption and enhancing comfort; additionally, the vehicle should arrive at the intersection when the light is green and pass through it at a high speed to avoid startup loss and improve the intersection efficiency. To meet these requirements, the trajectory optimization model proposed in~\cite{Zhou2017} and~\cite{Li2018} is introduced to optimize the vehicle trajectories in the coordinating area. \par

Given the length $L$ of the coordinating area and the entry time $t_i$, we first determine $t_o$, which is the time when the vehicle leaves the coordinating area and enters the intersection. Clearly, $t_o$ cannot be smaller than $t_i + d_{min}$, where $d_{min}$ is the minimum passing time of the area constrained by the entry velocity $\overline{v}$, the maximum allowance velocity in the conflict area $v_o$, and the acceleration boundaries $\underline{a}$ and $\overline{a}$, where $\underline{a} < 0 < \overline{a}$. In addition, some factors may further limit the feasible interval of $t_o$. The safety concern, which forces two adjacent vehicles to maintain a spatial gap, is the major limitation in this case. Meanwhile, vehicles cannot enter the intersection in red phases, so $t_o$ must be the earliest feasible time during a green phase.

\begin{figure}[!t]
    \centering
    \includegraphics[width=0.45\textwidth]{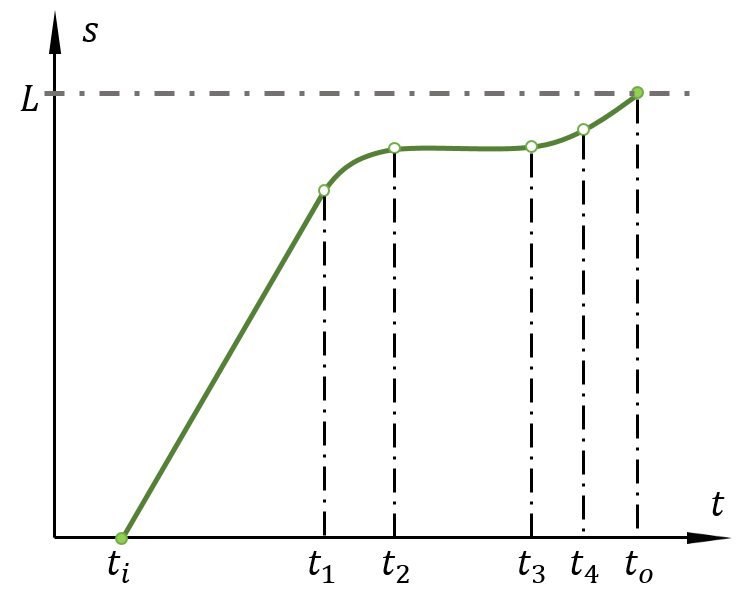}
    \caption{Five quadratic segments of a vehicle trajectory}
    \label{Fig2}
\end{figure}

\begin{figure}[!t]
	\centering
	\subfloat[][]{\includegraphics[width=0.4\textwidth]{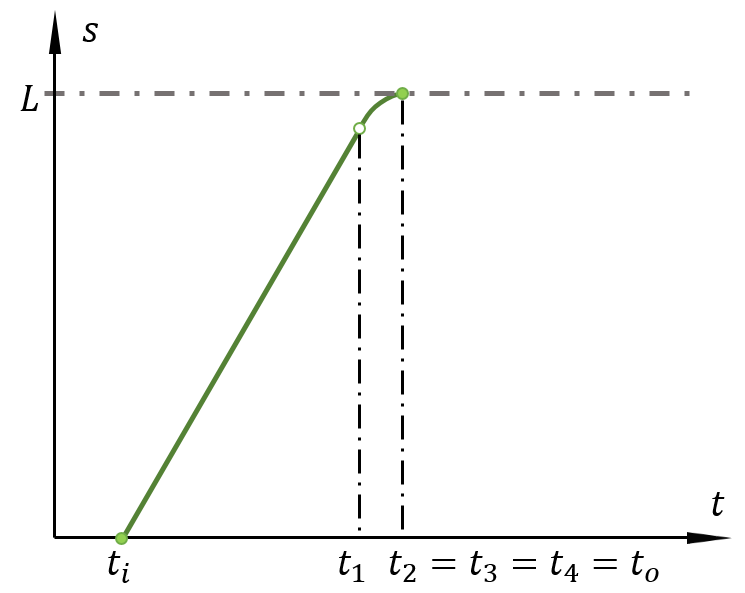}}\\
	\subfloat[][]{\includegraphics[width=0.4\textwidth]{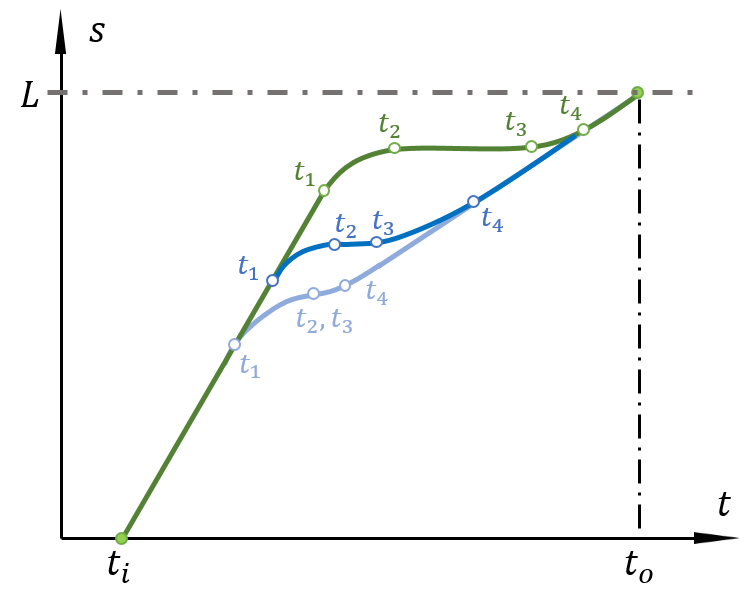}}
	\caption[]{Some possible trajectories with the same $t_i$ and $t_o$}
	\label{Fig3}
\end{figure}

Given $t_i$ and $t_o$, the determination of the entire vehicle trajectory is still difficult because this determination is an infinite-dimension problem. To simplify the planning process, all of the vehicles are arranged for a trajectory with five quadratic segments according to the method proposed in~\cite{Ma2017}. As illustrated in Fig. \ref{Fig2}, $t_1 \leq t_2 \leq t_3 \leq t_4 \in {[t_i,t_o]}$ denote the joint moments between segments. Vehicles first cruise at the entrance speed $\overline{v}$ in time interval $[t_i,t_1]$ and then decelerate at a constant deceleration rate $\underline{a}$ during $(t_1,t_2]$. In some cases, vehicles have to stop completely at $t_2$, and the length of $(t_2,t_3]$ denotes the duration that vehicles must remain stationary. Otherwise, $t_2$ is equal to $t_3$, and the third segment does not exist. Then, during the next segment starting at $t_3$, vehicles accelerate at $\overline{a}$ until their velocities reach the leaving speed $v_o$ at $t_4$. In the fifth segment during $(t_4,t_o]$, vehicles cruise at $v_o$ and enter the intersection at $t_o$. \par

It is straightforward that when the trajectory satisfies $t_2=t_3=t_4=t_o$, as shown in Fig. \ref{Fig3}(a), the travel time $d$ reaches its minimum and equals $d_{min}$. When $d = d_{min}$, the driving trajectory has a unique solution. However, when $d>d_{min}$, the trajectory cannot be uniquely determined. Fig. \ref{Fig3}(b) illustrates some feasible trajectories with the same $t_i$ and $t_o$. \par

\begin{figure*}[!t]
	\centering
	\subfloat[][]{\includegraphics[width=0.395\textwidth]{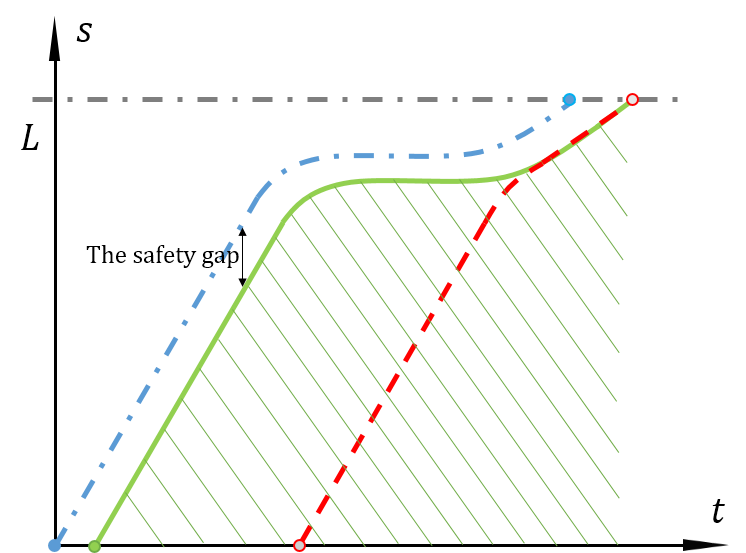}}
	\subfloat[][]{\includegraphics[width=0.515\textwidth]{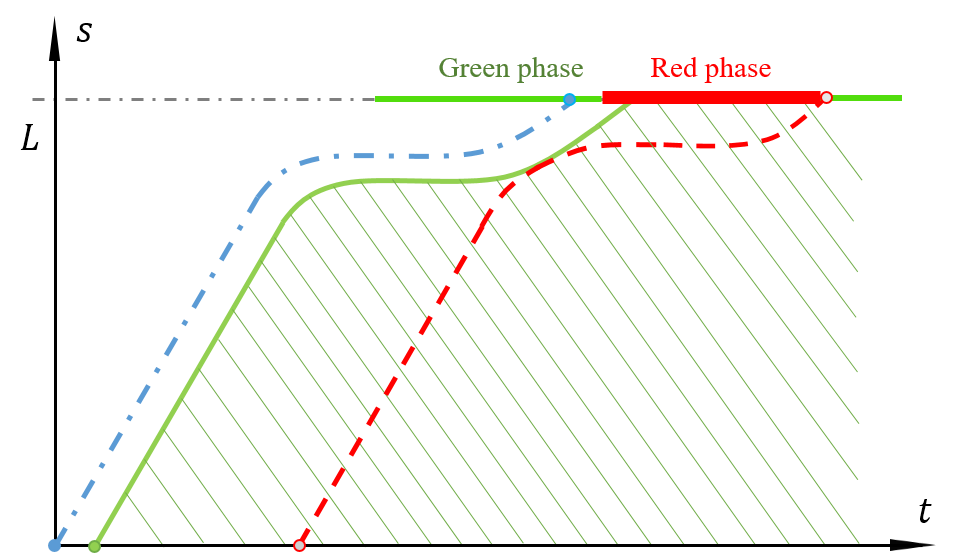}}
	\caption[]{Feasible region of the vehicle trajectory under the constraints of the preceding vehicle and signal lights}
	\label{Fig4}
\end{figure*}

Considering the safety constraint of the preceding vehicle in the same lane, the feasible region of the trajectory is further reduced. As illustrated in Fig. \ref{Fig4}, the upper boundary of the feasible trajectories is shown by the solid line, in which the distance from the preceding vehicle (for which the trajectory is shown in the dot dash line) at any time is equal to the minimum gap. Intuitively, in the selected trajectory, the deceleration time $t_1$ should be pushed back as much as possible to leave more viable space for the following vehicles. Thus, the optimal trajectory will be the one that is tangent to or coincident with the upper boundary at some point. In Fig. \ref{Fig4}(a), the selected trajectory is shown as the dashed line. Fig. \ref{Fig4}(b) adds the constraint of signal phases, delaying the entry of the vehicle to the intersection.

The calculation of the optimal trajectory exploits the properties of the quadratic function, which is cumbersome but not difficult. The optimal trajectory may be tangent to the upper boundary in the third segment or coincident with the boundary in the fourth and/or the fifth segment. A detailed of calculation to generate the trajectory is omitted in this paper.

\subsection{First-come-first-serve control} \label{method_FCFS_subsec}

For the ad-hoc negotiation based intersection control strategy, the \textit{first-come-first-serve} (FCFS) control~\cite{Dresner2008} is one of the most seminal works in the literature. The core idea of this control is to divide the intersection area into multiple tiles, so that the occupation of road space can be modeled as the occupation of tiles, reducing the variables used to describe the state of the intersection. In this control model, the safety gap is represented as an expanded area around vehicles, and the determination of occupied tiles is based on the expanded vehicle size. \par

\begin{figure*}[!t]
    \centering
    \includegraphics[width=0.7\textwidth]{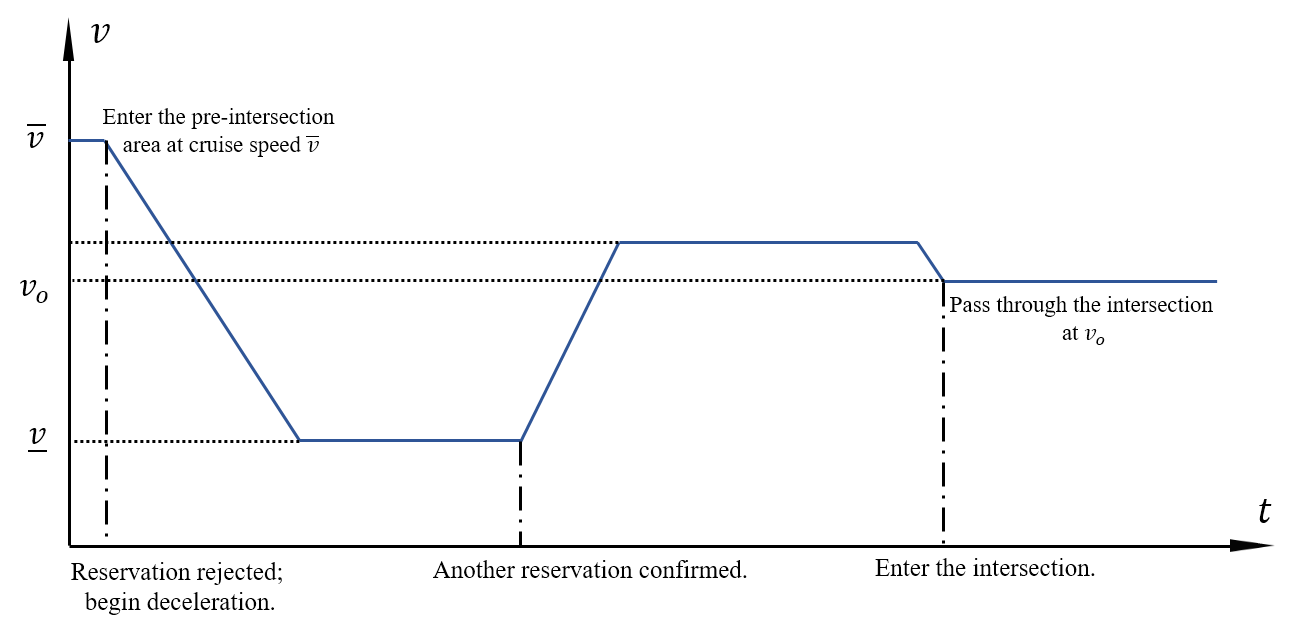}
    \caption{Possible speed-time curve of a vehicle under the FCFS strategy}
    \label{Fig5}
\end{figure*}

Whenever a CAV enters the coordinating area, the roadside traffic manager will try to reserve a feasible trajectory for it in which the occupied tiles do not coincide with the tiles reserved for previous vehicles at any moment. To avoid excessive calculation, the original model in~\cite{Dresner2008} only tested two trajectories each time for each reservation. The first trajectory allows vehicles to pass through the control section at the highest speed, which is $\overline{v}$ in the coordinating area and $v_o$ at the intersection. This trajectory causes no delay for the vehicle. The other trajectory guides the vehicle to pass through the intersection at its current speed. However, this trajectory may be inefficient because, in some cases when vehicles move at a low speed, the intersection is occupied for a longer time. To address this issue, we adjust the setting of the second trajectory to guarantee a high speed while passing through the intersection. When the first trajectory is revealed to be unfeasible, we test the feasibility of a trajectory for which vehicle pass through the coordinating area at a slightly lower speed, for example, $2\overline{v}/3$. The deceleration pushes back the time of the entry to the intersection, enabling the vehicles to pass through the intersection at the maximum allowable speed $v_o$ without conflicts. If both trajectories fail to be feasible, the vehicle will decelerate at the maximum rate $\underline{a}$. We allow the vehicles that do not have confirmed reservations to drive through the coordinating area at a safe speed $\underline{v}$ until they are stopped by queuing vehicles in front of the intersection or finally have their reservations confirmed. All the vehicles without confirmed reservations will send new requests at a certain frequency, which is higher when vehicles are closer to the intersection and lower when the requests are less urgent. A possible speed-time curve of a vehicle is illustrated in Fig. \ref{Fig5}. \par

\subsection{Conflict point intersection control} \label{method_CPIC_subsec}

Different from the FCFS control, planning-based intersection control considers multiple vehicles simultaneously in a dynamic fashion. In this category of control, an optimization problem is usually formulated to determine the intersection passing strategies for vehicles, with the objective function of minimizing delay or fuel consumption, and constraints such as speed limits and collision free. In addition, a rolling horizon model is generally adopted in dynamic traffic scenarios for real-time implementation~\cite{Meng2018}. Theoretically, planning-based control is considered to outperform ad-hoc negotiation based control because the strategies adopted in ad-hoc negotiation based control are always feasible solutions in planning-based control; however, constrained by the heavy computational burden in the optimization procedure, the performances of the planning-based control can only be sub-optimal in reality. \par

\begin{figure}[!t]
    \centering
    \includegraphics[width=0.45\textwidth]{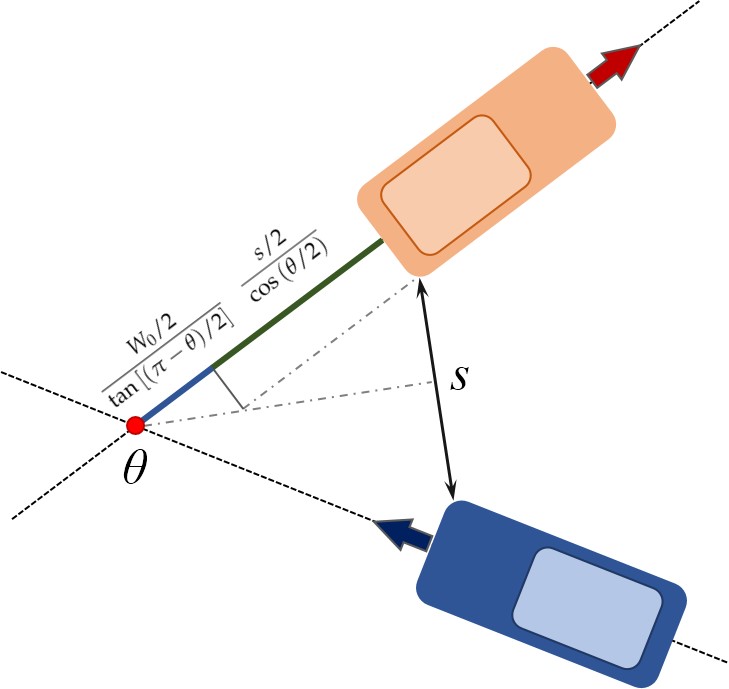}
    \caption{Determination of the safety buffer size}
    \label{Fig6}
\end{figure}

Our realization of planning-based control is mainly based on the Conflict Point Intersection Control (CPIC) model proposed in~\cite{Levin2017Confilct}. The conflict-point based collision-avoidance method was previously developed for aircraft management in airports~\cite{Liang2018} and the open air~\cite{Rey2016}; in the field of intersection control, Levin and Rey~\cite{Levin2017Confilct} define conflict points as locations where the vehicle trajectories of vehicles traveling from different directions intersect. The CPIC model then formulates a MILP problem to optimize the trajectories for passing vehicles. Under the assumption that vehicles maintain constant speed at intersections, we only need to optimize two parameters for each vehicle, namely, $t_i(r_i^-)$ and $t_i(r_i^+)$, which denote the intersection entry and exit time of vehicle $i$, respectively.  \par

To maintain the consistency of the comparison settings, the dynamic buffer size that varies with the vehicle speed in the original model~\cite{Levin2017Confilct} is replaced by a static safety gap in this paper. To guarantee the minimum gap, an extra longitudinal buffer $s'$ is added to each vehicle. As illustrated in Fig. \ref{Fig6}, given the safety gap $s$ between vehicles and the angle $\theta$ of the trajectories of vehicle $i,j$ on conflict point $c$, the buffer size $s'$ can be derived as

\begin{equation}
    s' = \frac{s/2}{\cos{(\theta/2)}}+\frac{W_0/2}{\tan{[(\pi-\theta)/2]}}
    \label{eq1}
\end{equation}

\noindent where $W_0$ is the width of vehicle $i$. It is noted that Eq.(\ref{eq1}) also holds when vehicles $i,j$ share the same trajectory, in which case $\theta$ equals to zero. The formation of the problem is as follows.\footnote{For conciseness, the problem formation is only briefly presented here. Readers may refer to~\cite{Levin2017Confilct} for more detailed description. }  \par

To begin, the objective function is:

\begin{equation}
    \min \sum_{i}t_i(r_i^+)
    \label{eq2}
\end{equation}

\noindent which minimizes the sum of exit time of all vehicles. Since the occurrence time is fixed for a given vehicle, the objective function also minimizes total delay. Eqs.(\ref{eq3})-(\ref{eq9}) are the constraints, including the speed limits, the first-in-first-out (FIFO) conditions and the collision-avoidance conditions. Firstly, we have \par

\begin{equation}
    t_i(r_i^-) \geq e_i
    \label{eq3}
\end{equation}

The constraint limits the earliest intersection entry time. \par

\begin{equation}
    t_i(r_i^-) + \tau_i(r_i^-) \leq t_j(r_j^-)
    \label{eq4}
\end{equation}

\begin{equation}
    t_i(r_i^+) + \tau_i(r_i^+) \leq t_j(r_j^+)
    \label{eq5}
\end{equation}

Eqs.(\ref{eq4}) and (\ref{eq5}) guarantee the FIFO constraint for all vehicles $i$,$j$ that have the same spatial trajectories, \noindent where

\begin{equation}
    \tau_i(c) = \frac{L_i(c) \cdot (t_i(r_i^+) - t_i(r_i^-))}{d(r_i^-,r_i^+)}
    \label{eq6}
\end{equation}\par

 \noindent denotes the time duration that vehicle $i$ occupies conflict point $c$. In Eq.(\ref{eq6}), $L_i(c)$ denotes the vehicle length with the safety buffer included. Therefore, we have $L_i=L_0+2s'$, where $L_0$ is the physical length of the vehicle, and $s'$ is the safety buffer required for conflict point $c$.

\begin{equation}
    \frac{d_i(r_i^-,r_i^+)}{\overline{U_i}} \leq t_i(r_i^+) - t_i(r_i^+) \leq \frac{d_i(r_i^-,r_i^+)}{\underline{U_i}}
    \label{eq7}
\end{equation}

Eq.(\ref{eq7}) provides the upper ($\overline{U_i}$) and lower ($\underline{U_i}$) speed limits for vehicles. For every pair of vehicles $i$,$j$ that come from different directions and for which their trajectories intersect at the conflict point $c$, we have \par

\begin{equation}
    t_i(c) + \tau_i(c) - t_j(c) \leq (1 - \delta_{ij}(c))M_{ij}
    \label{eq8}
\end{equation}

\noindent where $M_{ij}$ is a large number, and $\delta_{ij}(c)$ is a binary variable representing the passing order of vehicles $i$,$j$ at point $c$. If $j$ enters $c$ after $i$ has left, then $\delta_{ij}(c)=1$ and $\delta_{ji}(c)=0$, and vice versa. Therefore, we have \par

\begin{equation}
    \delta_{ij}(c) + \delta_{ji}(c) = 1
    \label{eq9}
\end{equation}

Eqs.(\ref{eq2})-(\ref{eq9}) formulate the MILP problem involved in the CPIC model. Due to the numerous binary variables $\delta_{ij}$, the number of which is twice that of vehicle pairs that have conflict points in their paths, obtaining a solution of this problem is quite time-consuming. According to the simulation experiment conducted in~\cite{Levin2017Confilct}, the trajectory arrangement for no more than 30 vehicles can be completed in real time. Therefore, the rolling horizon framework is also adopted in the CPIC to limit the vehicle number, ensuring that the model is feasible for use in a real traffic system. \par

\begin{figure}[!t]
    \centering
    \includegraphics[width=0.45\textwidth]{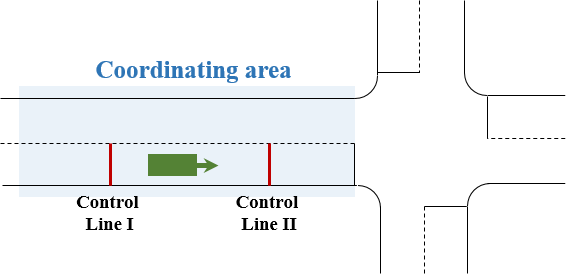}
    \caption{Two control lines in the rolling horizon framework}
    \label{Fig7}
\end{figure}

The rolling horizon framework is described as follows. As illustrated in Fig. \ref{Fig7}, the framework includes two control lines: the outer line is regarded as “vision”, and the inner line suggests the minimum safety distance for a vehicle to adjust its speed to pass through the intersection. The intersection manager receives or detects the location and the velocity of all approaching CAVs at every time step, and when a vehicle passes through the first control line, it will be added into the optimization set. In each time step, the traffic manager updates the members of the optimization set, and then, the MILP problem is formulated to find the optimal trajectories for vehicles. Presented as the “as late as possible” (ALAP) rule in~\cite{Levin2017Confilct}, the trajectory for each vehicle remains adjustable until it passes the second control line. The passed vehicles remain in the optimization set, providing constraints on the feasible trajectories of other vehicles until they have traveled so far that cannot have any impact on the vehicles for which the trajectories have not yet been determined. \par

While the position of the inner control line is determined by the speed limit of the coordinating area and the intersection, the position of the outer line should be carefully analyzed. If the gap between the two lines is set to a too large value, too many vehicles will be involved in the optimization, and therefore, real-time trajectory allocation will be impossible; on the other hand, the performance of the solution will be damaged by a narrow gap. In the simulations of this paper, the outer control line is set as far from the intersection as possible, on the condition that the optimization can be completed soon enough.  \par

\section{Comparisons of numerical simulations} \label{simulation_sec}

In this section, we present the results of the numerical simulations on the three control models under a variety of scenarios. By using heterogeneous intersection scenarios, we assess the model performance characteristics under different traffic demand patterns and intersection layouts. We first test the model performance characteristics in a symmetric 4-leg intersection (scenarios \textbf{1-A} and \textbf{1-B}). Then, in scenarios \textbf{2-A} and \textbf{2-B}, two legs from the opposite direction were narrowed as secondary roads. Similar to the 4-leg intersection, scenarios \textbf{3-A} and \textbf{3-B} test the performance in a T-type junction. In each intersection layout, we ran the simulation under both balanced and imbalanced traffic demand. Moreover, we explored the possible impact of the fluctuating vehicle arrival sequences (in scenarios \textbf{4-A} and \textbf{4-B}) and different safety buffers (in scenarios \textbf{5-A} and \textbf{5-B}). The numerical simulations were coded on MatLAB, and conducted using a personal computer with an AMD Ryzen 5 2600 CPU and 16GB RAM. \par

For an unbiased comparison of the performance of the vehicle-based and phase-based traffic control, we ensure that each control strategy shares exactly the same traffic scenario and environmental variables. The time step as well as the reaction time in simulations is 0.2 seconds. The length of the coordinating area is 600 m. Vehicles enter the area at the maximum allowable cruising speed ($\overline{v}$), which is 18 m/s in simulations; the minimum speed $\underline{v}$ along the coordinating area is set 5 m/s. When passing through the intersection, the speed limit $v_o$ is 15 m/s for through vehicles and 10 m/s for left-turn vehicles, while right turns are ignored from the model due to their negligible influence on the intersection traffic. The maximum acceleration and deceleration rates are 1.5 $m/s^2$. In the phase-based traffic control, the phase transition loss between two consecutive phases is 3 seconds. All of the vehicles are cars with dimensions of 4 m $\times$ 1.8 m. To simplify the model, we adopt a static buffer size that does not vary with vehicle speed. In scenarios \textbf{1}, \textbf{2}, \textbf{3} and \textbf{4}, the minimum spatial gap between any two vehicles is 1.0 m, and we also test the performance of the control models under different safety gap settings of 4.0 and 8.0 m in scenarios \textbf{5-A} and \textbf{5-B}.

\subsection{Traffic generation} \label{simulation_TraffGen_subsec}

In the simulations, we use $\lambda_0$ to describe the traffic demand volume, which denotes the average number of total arrivals per second in all lanes. In each scenario, $\lambda_0$ varies from 0.1 to 4.0, representing the total traffic volume from 360 to 14,440 vehicles per hour (vph). In addition to the total volume, we also specify the distribution pattern of traffic demands by setting the vehicle distribution ratio $r_i$ on lane $i$, where $\sum_ir_i=1$. The vehicle arrival rate of lane $i$ is determined by the following Eq.(\ref{eq10}). \par

\begin{equation}
    \lambda_i=\lambda_0r_i
    \label{eq10}
\end{equation}

For each vehicle arrival vector \textbf{$\lambda$}, 10 realizations of vehicles arrivals in 65 minutes are randomly generated. As suggested in~\cite{Korkmaz2010}, we supposed that the headway follows a shifted exponential distribution, and the minimum following headway is set to 1.0 s. To approximately describe the arrival patterns, we discretize the time and set each time step as 0.2 s; in each time step, the probability that a new arrival is generated, i.e., $p_i$, is determined by Eq.(\ref{eq11}) (as long as the time gap to the previous vehicle is no less than 1.0 s). By Eq.(\ref{eq11}), vehicles are generated following an approximated Poisson distribution with a minimum gap of 1.0 s.

\begin{equation}
    p_i = \frac{0.2\lambda_i}{1-\lambda_i}
    \label{eq11}
\end{equation}

With the generated arrival realizations, the three control models are tested, and the differences between the simulated travel times and the free flow times of all of the vehicles are aggregated as the total delay. Vehicle delays during the first 5 min are not counted in the simulation. In some extreme cases, the queuing vehicles may spill back to the head of the coordinating area and therefore block new arrivals. If the generation of vehicles is blocked, the entrance will be postponed and the waiting time is also considered into the total delay. \par

\begin{figure}[!t]
	\centering
	\subfloat[][]{\includegraphics[width=0.35\textwidth]{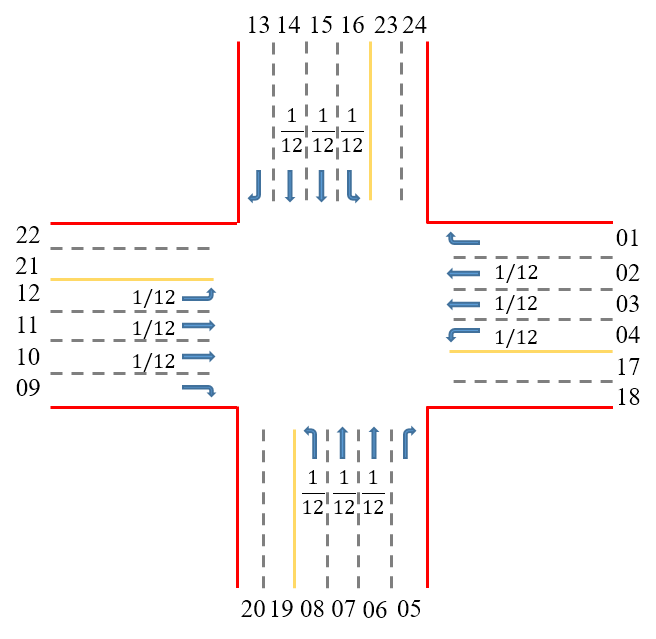}}\\
	\subfloat[][]{\includegraphics[width=0.35\textwidth]{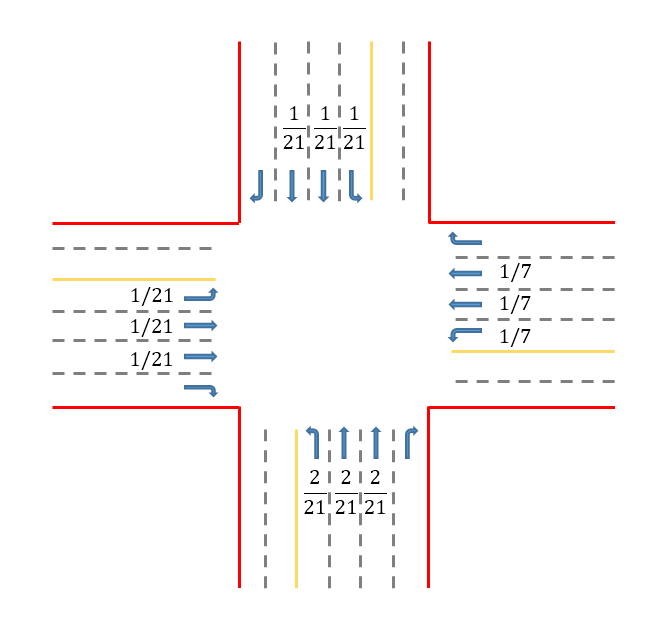}}
	\caption[]{Intersection layouts and demand patterns under scenarios \textbf{1-A}(a) and \textbf{1-B}(b). The numbers on the lane represent the demand level of the lane.}
	\label{Fig8}
\end{figure}

\begin{figure}[!t]
	\centering
	\subfloat[][]{\includegraphics[width=0.45\textwidth]{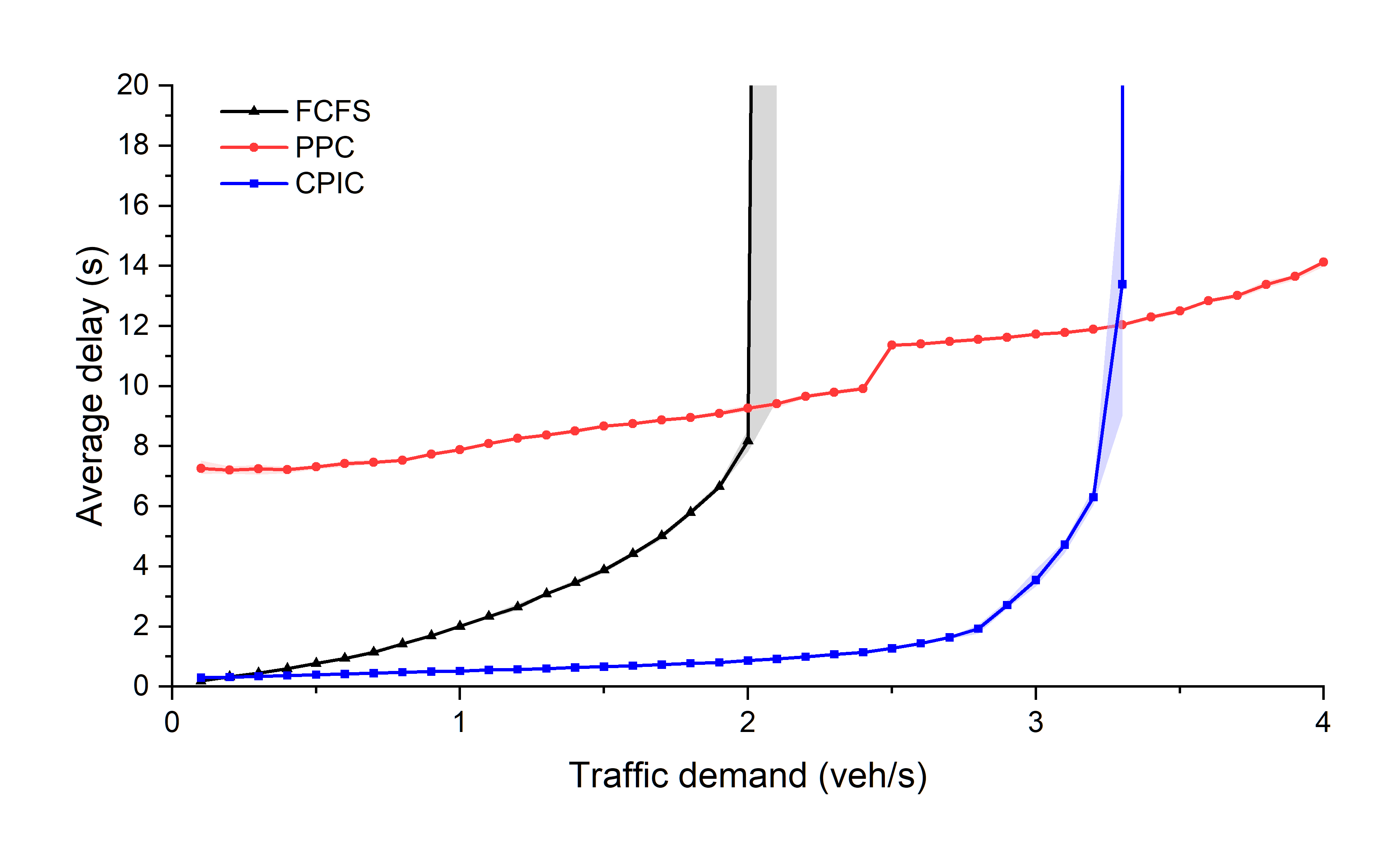}}\\
	\subfloat[][]{\includegraphics[width=0.45\textwidth]{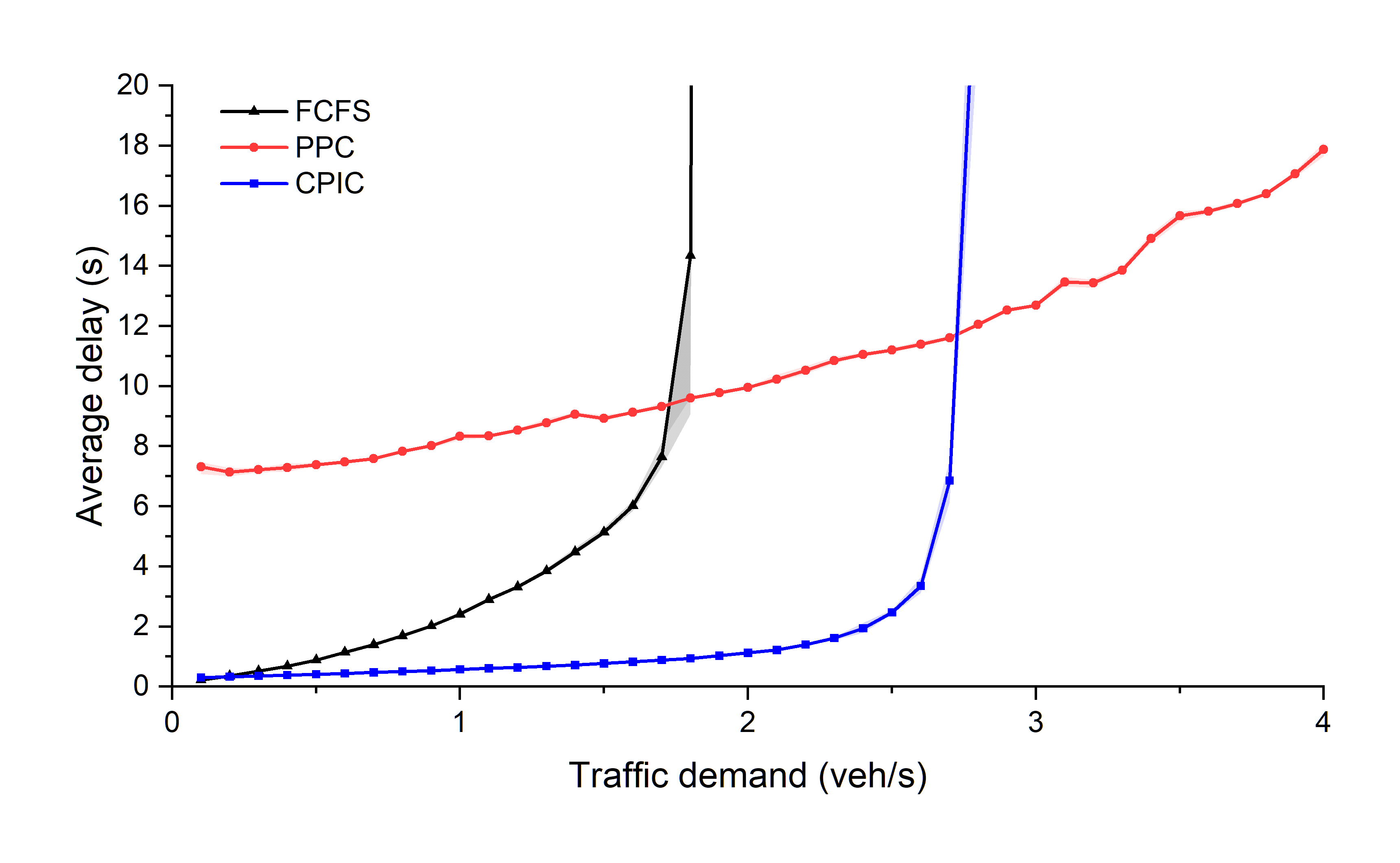}}
	\caption[]{Simulation results of scenarios \textbf{1-A}(a) and \textbf{1-B}(b)}
	\label{Fig9}
\end{figure}

\subsection{Symmetric 4-leg intersection} \label{simulation_Sce1_subsec}

As shown in Fig. \ref{Fig8}, we model a 4-leg intersection with 6 lanes in each leg, among which 4 lanes are approaching the intersection and 2 lanes are departing. The lane width is set to 3 m. As illustrated, the lanes are numbered from 1 to 24. The possible routes between the lanes are fixed; for example, a vehicle that arrives from the south and intends to make a left turn heading west will enters the intersection from lane 12 and leaves at lane 23. The right-turn movements are omitted in this study because they have no direct conflict with other movements, indicating that no vehicle is generated from lanes 1, 5, 9 and 13. Two traffic demand patterns are tested under this intersection structure: the balanced demand pattern and the imbalanced demand pattern. \par

In the balanced demand pattern (scenario \textbf{1-A}), the arrival rates of all of the lanes are equally set as $1/12$ of the total arrival rate $\lambda_0$. Therefore, we have $r_i=1/12$ for $i \in \{2,3,4,6,7,8,10,11,12,14,15,16\}$ and $r_i=0$ otherwise, as shown in Fig. \ref{Fig8}(a). For every demand level $\lambda_0$ from 0.1 to 4.0, Fig. \ref{Fig9}(a) presents the average value of the delay time of 10 experiments, and the 25th and 75th values are expressed by the colored region around the curve. The comparisons across various intersection control models (ad-hoc negotiation based FCFS, planning-based CPIC, and phase-based PPC strategy) show notable differences in the average delay, particularly when traffic demand is high. \par

The simulation results of the PPC strategy show a remarkable improvement in traffic efficiency. Benefiting from the reduced headway and start-up losses, the cycle length of the improved control can be greatly shortened, decreasing the average delay of passing vehicles. When traffic demand is quite low, a green time proportion of 25\% is verified to be sufficient for cleaning queuing vehicles, and the average delay is therefore decreased to no more than 8 s. The two vehicle-based traffic control models show even better performance under low traffic volume. Most vehicles can maintain the maximum speed when passing the intersection, experiencing negligible delays. In most traffic demands, planning-based CPIC strategy outperforms ad-hoc negotiation based FCFS strategy, since the FCFS strategy is always within the feasible region of the CPIC problem. However, it may seem counter-intuitive that CPIC exhibits a slightly larger delay than FCFS under extremely low traffic demand. This is due to the approximation we impose to eliminate nonlinear constraints that forces the vehicles to slow down slightly while approaching the intersection. \par

As the traffic demand increases, the average delays incurred by the three control models all increase. For the PPC strategy, a higher proposition of green time is required, causing higher cycle lengths, consequently resulting in a higher average delay. However, the increase in the average delay is quite small compared to that of FCFS and CPIC strategies. Under the scenario that the traffic demand is 4.0 vehicles per second (or 14,400 vph), the numerical simulations show that the average delay is 14.131 s under an optimal cycle length of 30 s. On the other hand, as shown in Fig. \ref{Fig9}(a), the vehicle-based traffic control models show different curves. In the FCFS control, the average delay for $\lambda_0=2.0$ is 8.178 s, while under the demand that $\lambda_0=2.1$, the delay time increases to 118.890 s. At the end of the simulation, we observed a maximum delay of 262.130 s, indicating that the vehicle queue is constantly growing. Similar phenomena were also observed in the CPIC model when $\lambda_0$ reached 3.4. In this case, traffic demand has reached the capacity of the intersection under the control strategies. \par

For this intersection layout, we can briefly summarize the performance of the different intersection control models under various demand levels. The vehicle-based traffic control performs well under low demand, but the capacity of these models is relatively low, i.e., the models cannot accommodate large demands well. On the contrary, the phase-based traffic control shows higher delay under low traffic volume scenarios, but in high demand cases ($\lambda_0 \geq 3.3$ in this scenario), it becomes the only method that can stabilize the intersection queues. \par

In the imbalanced distribution pattern (scenario \textbf{1-B}), we assumed that the traffic demand from the east and the south is higher than average. Fig. \ref{Fig8}(b) illustrates the distribution: $r_i=1/21$ for $i \in \{10,11,12,14,15,16\}$, $r_i=2/21$ for $i \in \{2,3,4\}$ and $r_i=1/7$ for $i \in \{6,7,8\}$. The simulation results presented in Fig. \ref{Fig9}(b) lead to similar conclusions to those as in scenario \textbf{1-A}. Both vehicle-based traffic control models provide high-quality and stable intersection management under low traffic demand ($\lambda_0 \leq 1.9$ for FCFS and $\lambda_0 \leq 3.0$ for the CPIC strategy), but in high-demand scenarios, only the delay of the PPC strategy remains acceptable. \par

\subsection{4-leg intersections with secondary roads} \label{simulation_Sce2_subsec}

\begin{figure}[!t]
	\centering
	\subfloat[][]{\includegraphics[width=0.35\textwidth]{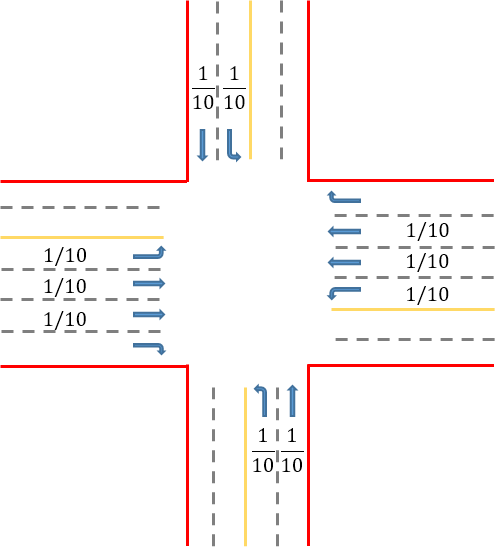}}\\
	\subfloat[][]{\includegraphics[width=0.35\textwidth]{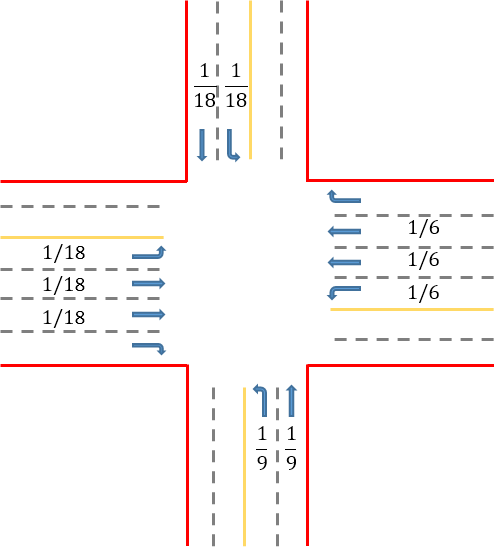}}
	\caption[]{Intersection layouts and demand distribution under the scenarios \textbf{2-A}(a) and \textbf{2-B}(b)}
	\label{Fig10}
\end{figure}

We then examine the performances of the control models under a smaller intersection where a 6-lane main road intersects with a 4-lane secondary road. Moreover, the balanced (scenario \textbf{2-A}) and imbalanced (scenario \textbf{2-B}) traffic patterns are included. The intersection layouts and the demand patterns are presented in Figs.\ref{Fig10}(a) and \ref{Fig10}(b), respectively. This type of intersections is prevalent in urban road networks, particularly on arterial
roads. \par

\begin{figure}[!t]
	\centering
	\subfloat[][]{\includegraphics[width=0.45\textwidth]{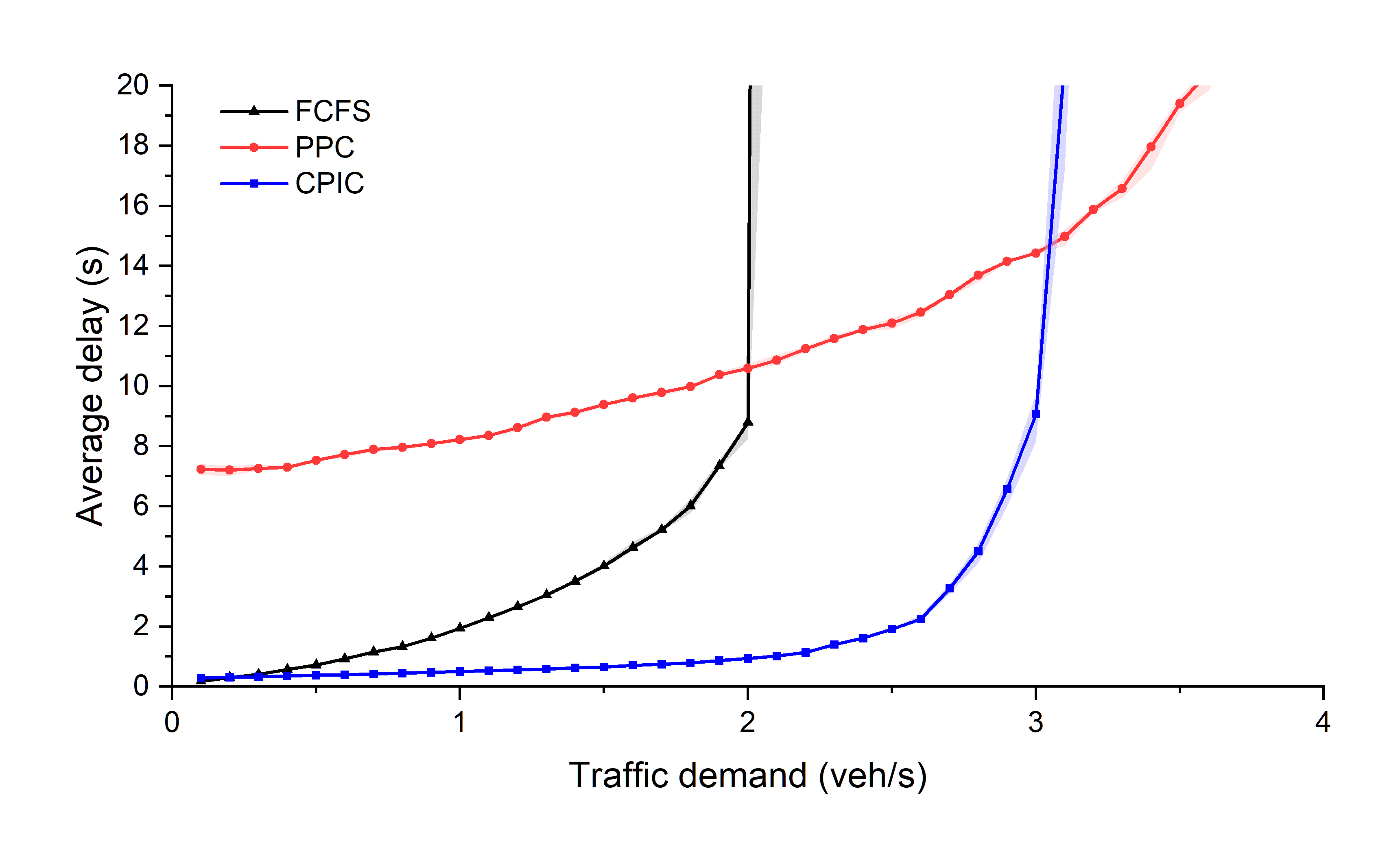}}\\
	\subfloat[][]{\includegraphics[width=0.45\textwidth]{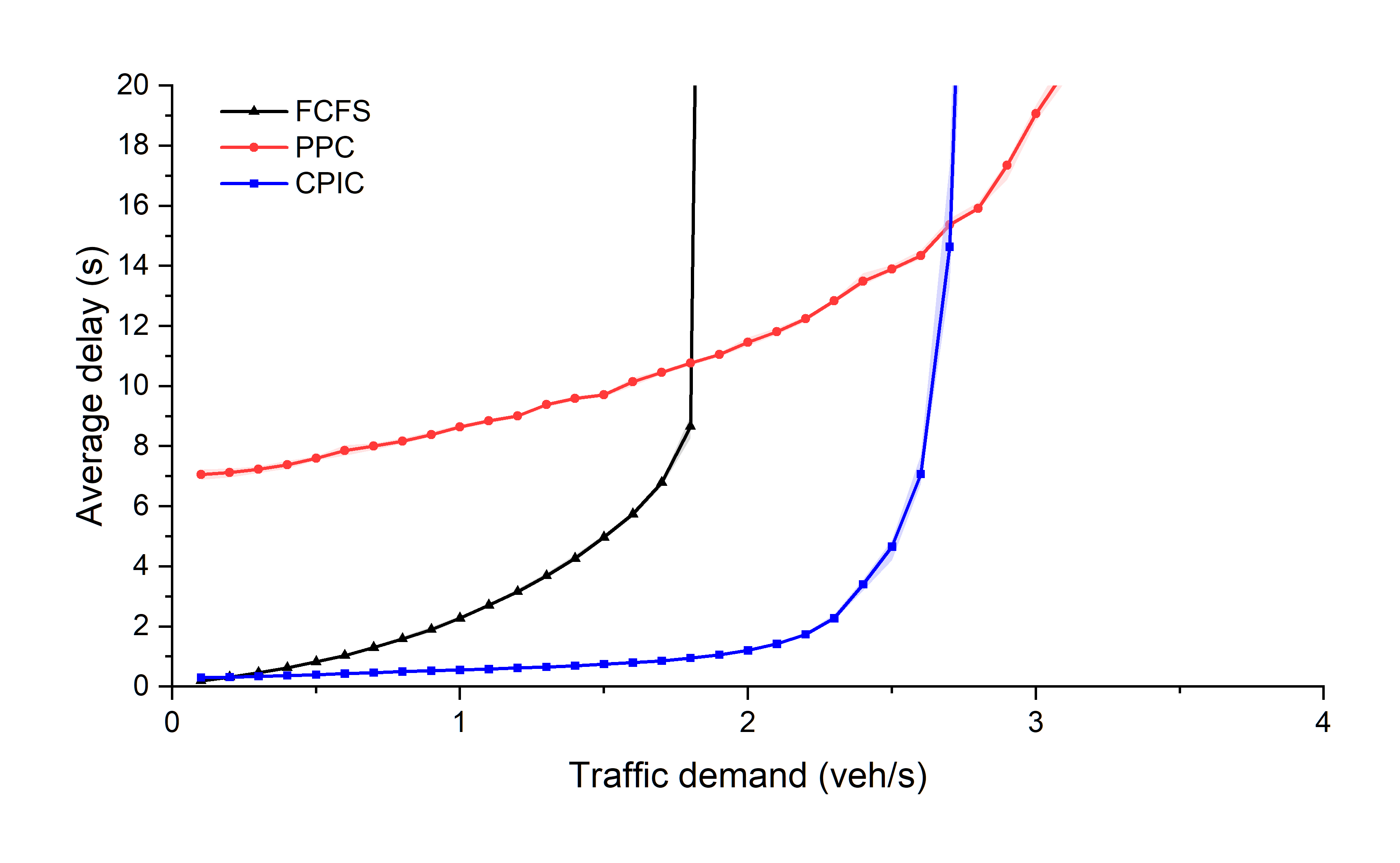}}
	\caption[]{Simulation results of scenarios \textbf{2-A}(a) and \textbf{2-B}(b)}
	\label{Fig11}
\end{figure}

The simulation results under the balanced and imbalanced traffic patterns are presented in Figs.\ref{Fig11}(a) and \ref{Fig11}(b), respectively. The basic trends of the average delay given by the three control models in this intersection layout do not differ much from those observed in scenarios \textbf{1-A} and \textbf{1-B}. It should be noted that in scenario \textbf{2-A}, a relatively high demand of $\lambda_0=1.8$ caused a significant deviation among the 10 simulations under the FCFS strategy. The results indicate that the performance of the FCFS strategy in busy intersections is unreliable with the potential risks for intersection failure. \par

\subsection{T-type junctions} \label{simulation_Sce3_subsec}

\begin{figure}[!t]
	\centering
	\subfloat[][]{\includegraphics[width=0.35\textwidth]{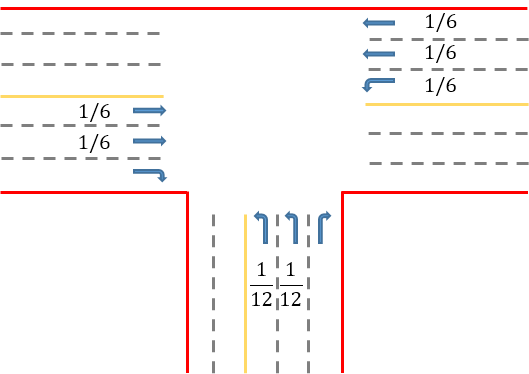}}\\
	\subfloat[][]{\includegraphics[width=0.35\textwidth]{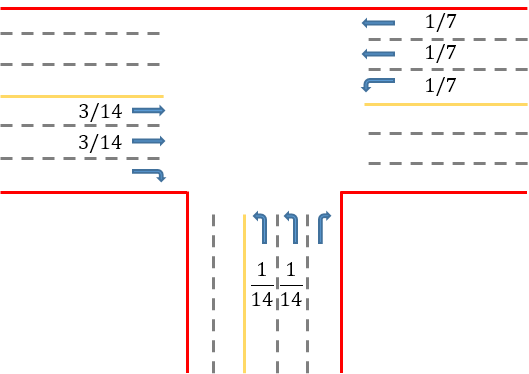}}
	\caption[]{Layouts and demand distribution of the T-type junction}
	\label{Fig12}
\end{figure}

A series of simulations were also used to examine T-type junctions, which are another important category of intersections. In the junction connecting one main road (from the west and east) and a secondary road (from the south), traffic from the main road is dominant. The intersection layouts and the demand distributions are presented in Figs.\ref{Fig12}(a) and \ref{Fig12}(b), including a balanced demand pattern and an imbalanced one. \par

\begin{figure}[!t]
	\centering
	\subfloat[][]{\includegraphics[width=0.45\textwidth]{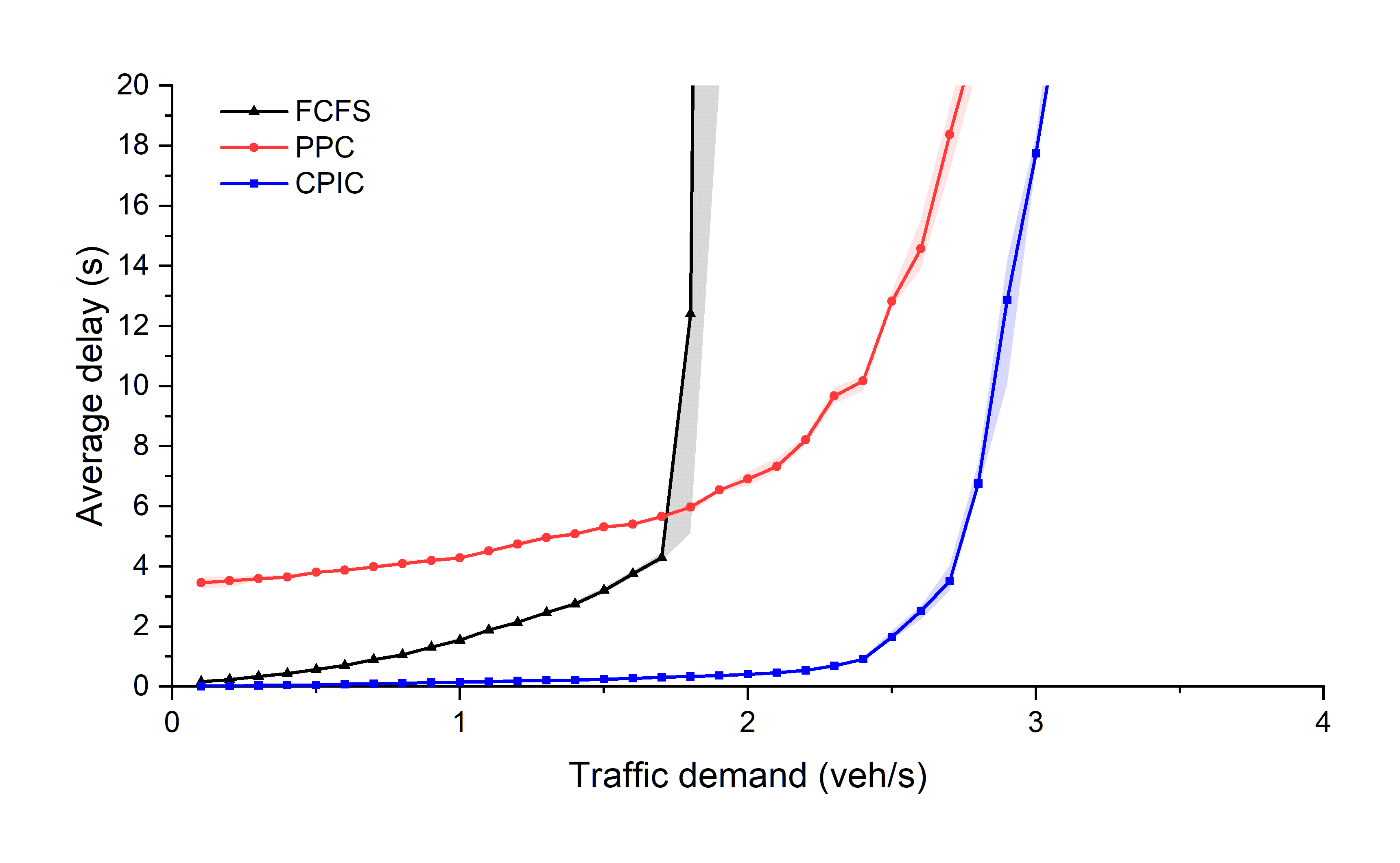}}\\
	\subfloat[][]{\includegraphics[width=0.45\textwidth]{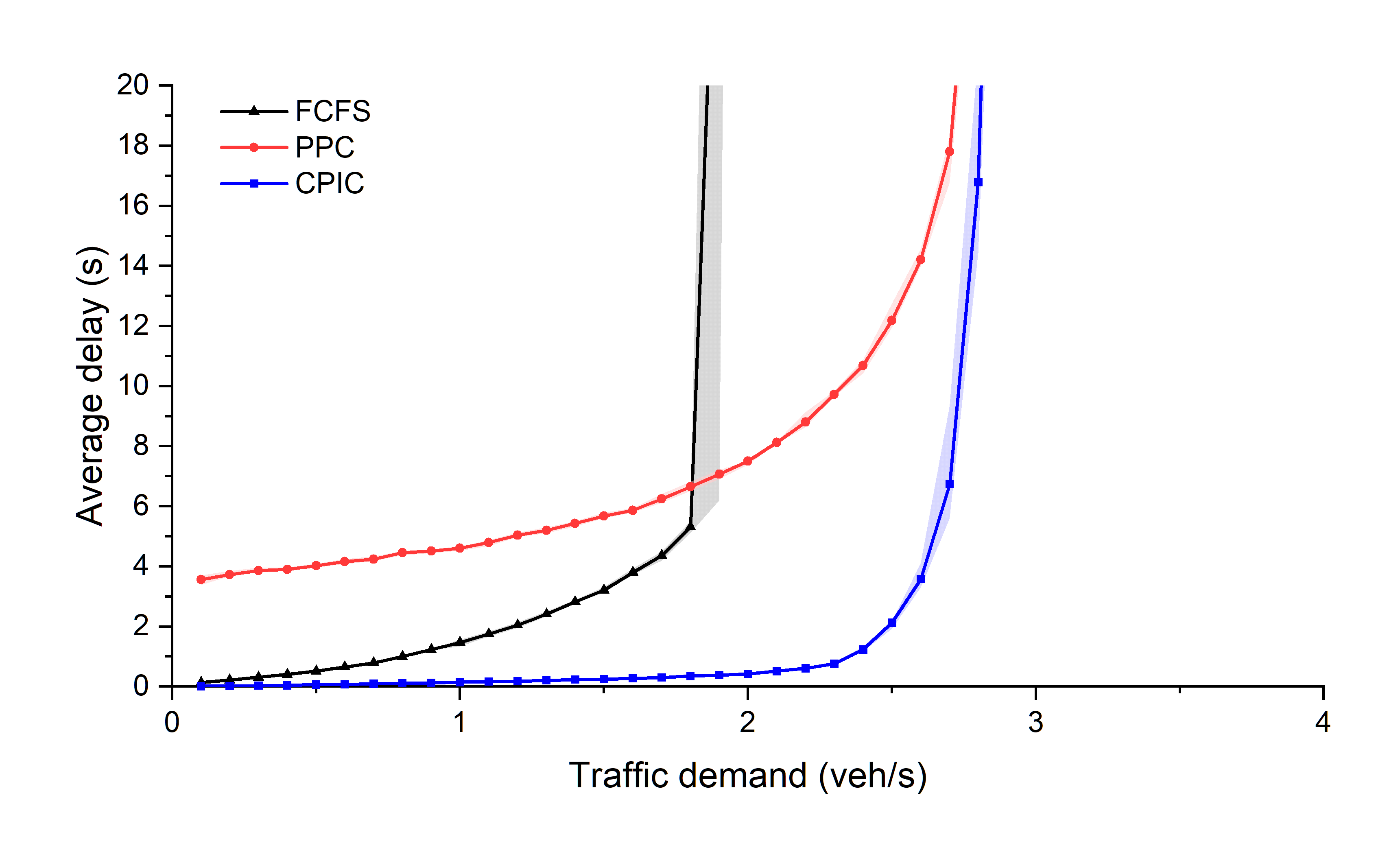}}
	\caption[]{Simulation results of scenarios \textbf{3-A}(a) and \textbf{3-B}(b)}
	\label{Fig13}
\end{figure}

\begin{figure}[!t]
	\centering
	\subfloat[][]{\includegraphics[width=0.4\textwidth]{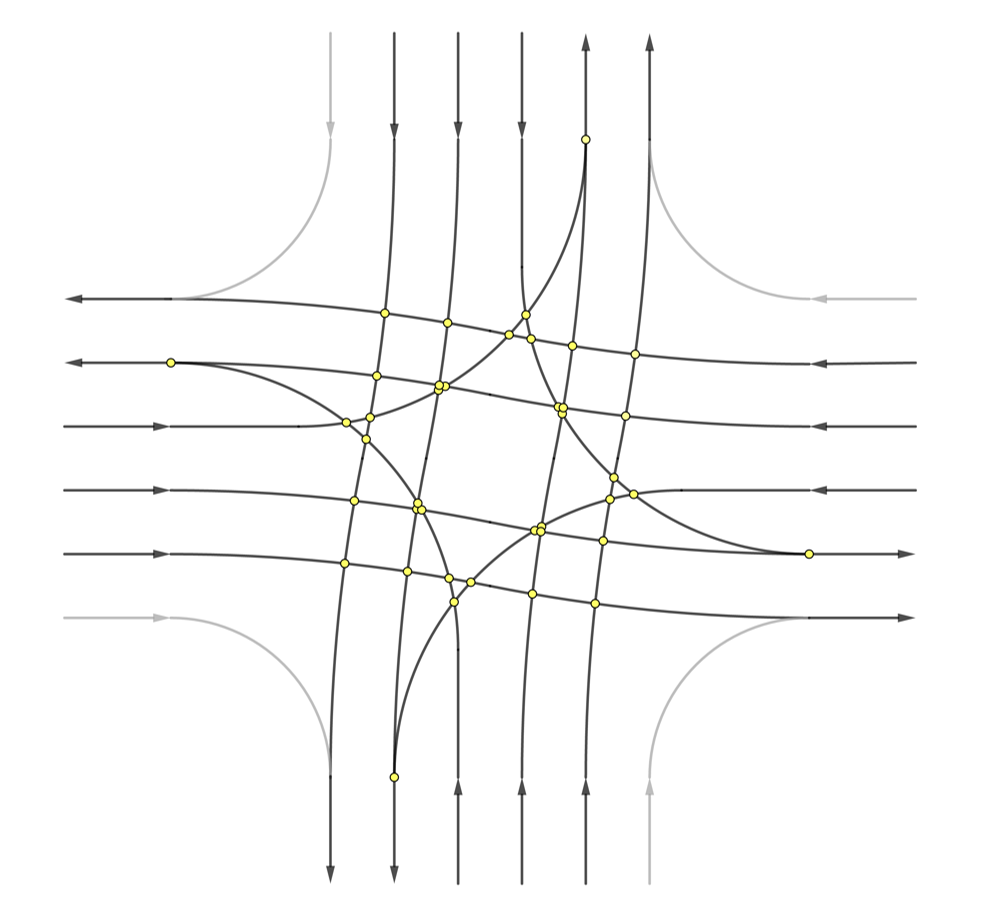}}\\
	\subfloat[][]{\includegraphics[width=0.4\textwidth]{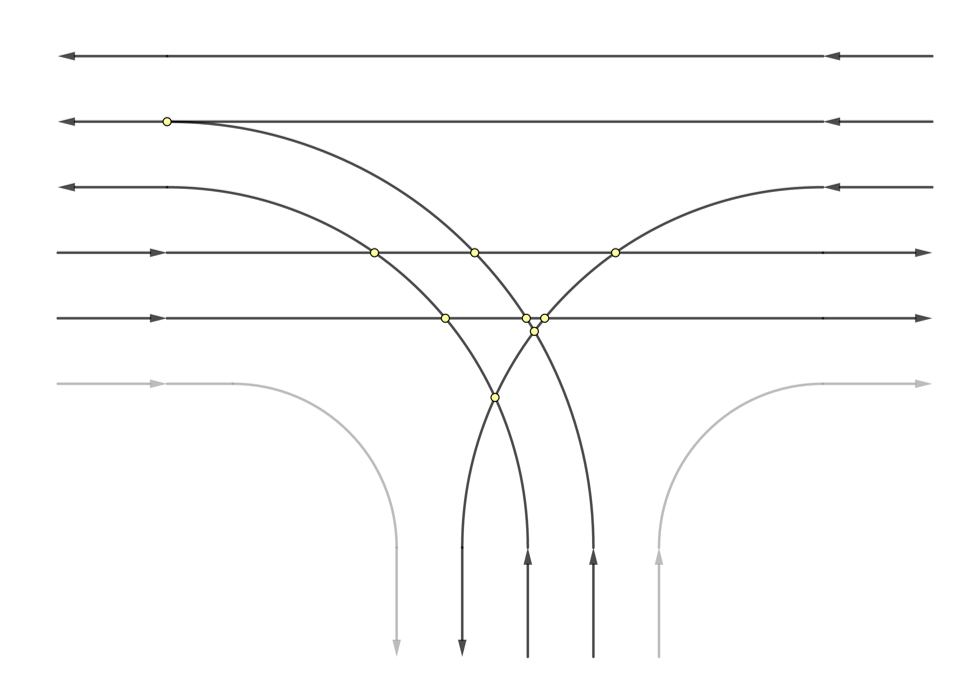}}
	\caption[]{Conflict points in 4-leg intersections (a) and T-type junctions (b)}
	\label{Fig14}
\end{figure}

As shown in Figs.\ref{Fig13}(a) and \ref{Fig13}(b), the simulation results under the T-type junction are quite different from those obtained for the 4-leg intersections. The planning-based CPIC strategy is revealed to show the best performance for all demand levels. This result may be due to the asymmetrical structure of the junction, which significantly affects the performance of the PPC strategy. In the T-type junction, two out of three phases are dedicated for left-turn movements from one specific direction, providing limited passing permissions for vehicles to go through the junction. Therefore, the capacity of the PPC strategy has experienced a sharp decline in simulations. On the contrary, the CPIC strategy benefits from the simplification of conflict relations. As illustrated in Figs. \ref{Fig14}, the number of conflict points (shown as circle dots) is 40 in 4-leg intersections, and 9 in T-type junction. Consider a scenario with 12 vehicles evenly distributed; the ratio of vehicles that arrive from each lane is shown in Fig. \ref{Fig8} for 4-leg intersections and Fig. \ref{Fig12} for T-type junctions. We can then calculate the total numbers of conflicting vehicle pairs that must be handled simultaneously in the CPIC strategy, which are 40 in 4-leg intersections and 24 in T-type junctions, respectively. In this case, consequently, the number of 0-1 variables in T-type junctions is 48, which is 32 less than that in 4-leg intersections. It notably lessens the computing burden of the optimizing problem in T-type junctions, and makes it possible to adopt a larger optimization horizon and acquire better solutions in scenarios with high traffic density. Hence, we observed the advantageous performances of the CPIC strategy compared to those of the PPC strategy in the T-type intersection. \par

\subsection{Symmetric 4-leg intersection under fluctuating arrival rates} \label{simulation_Sce4_subsec}

In following scenarios, we examine the impact of fluctuating arrivals in a symmetric 4-leg intersection. The intersection layouts and the demand distributions are the same as in scenarios \textbf{1-A} and \textbf{1-B}, which are shown in Figs.\ref{Fig8}(a) and \ref{Fig8}(b), respectively. The differences lie in the traffic generation procedure. In scenarios \textbf{4-A} and \textbf{4-B}, the vehicle generation rate varies every 2 minutes between $0.5\lambda_0$ and $1.5\lambda_0$. For instance, for the demand volume such that $\lambda_0=2.0$, the average vehicle arrival rate is set to 1.0 vehicle per second in the first two minutes of the simulation and 3.0 vehicles per second during the next two minutes. The purpose of these experiments is to study the ability of the control models to deal with temporary queues. In traditional traffic scenarios, joining a growing queue generally leads to an additional queuing delay. Considering the notable startup loss in manual driving, it will take even more time for a queue to dissipate, giving rise to a degradation in the intersection efficiency. However, due to the shorter reaction time of CAVs, it is expected that the fluctuating arrival process will lead to a weaker impact on autonomous driving intersections. \par

\begin{figure}[!t]
	\centering
	\subfloat[][]{\includegraphics[width=0.45\textwidth]{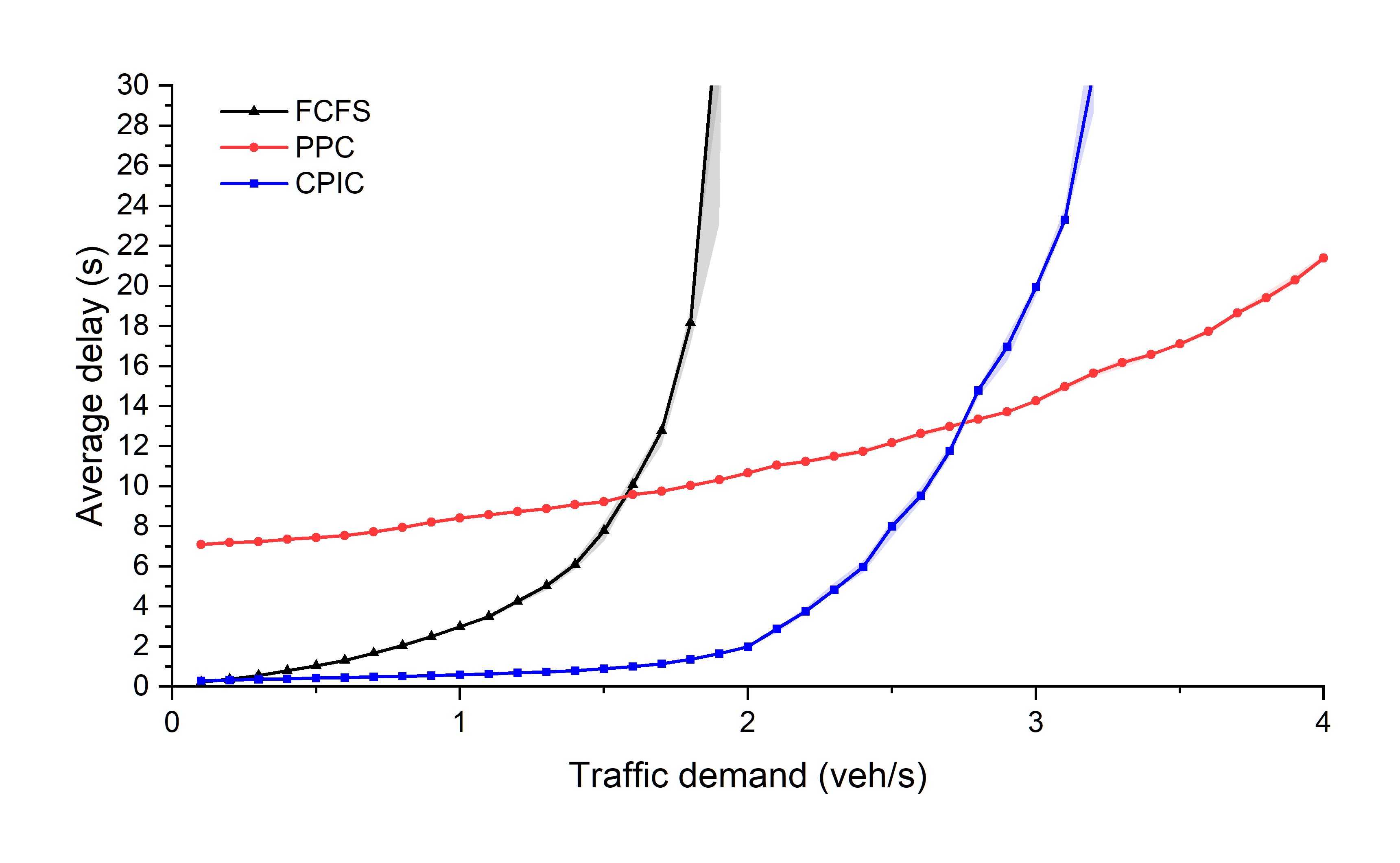}}\\
	\subfloat[][]{\includegraphics[width=0.45\textwidth]{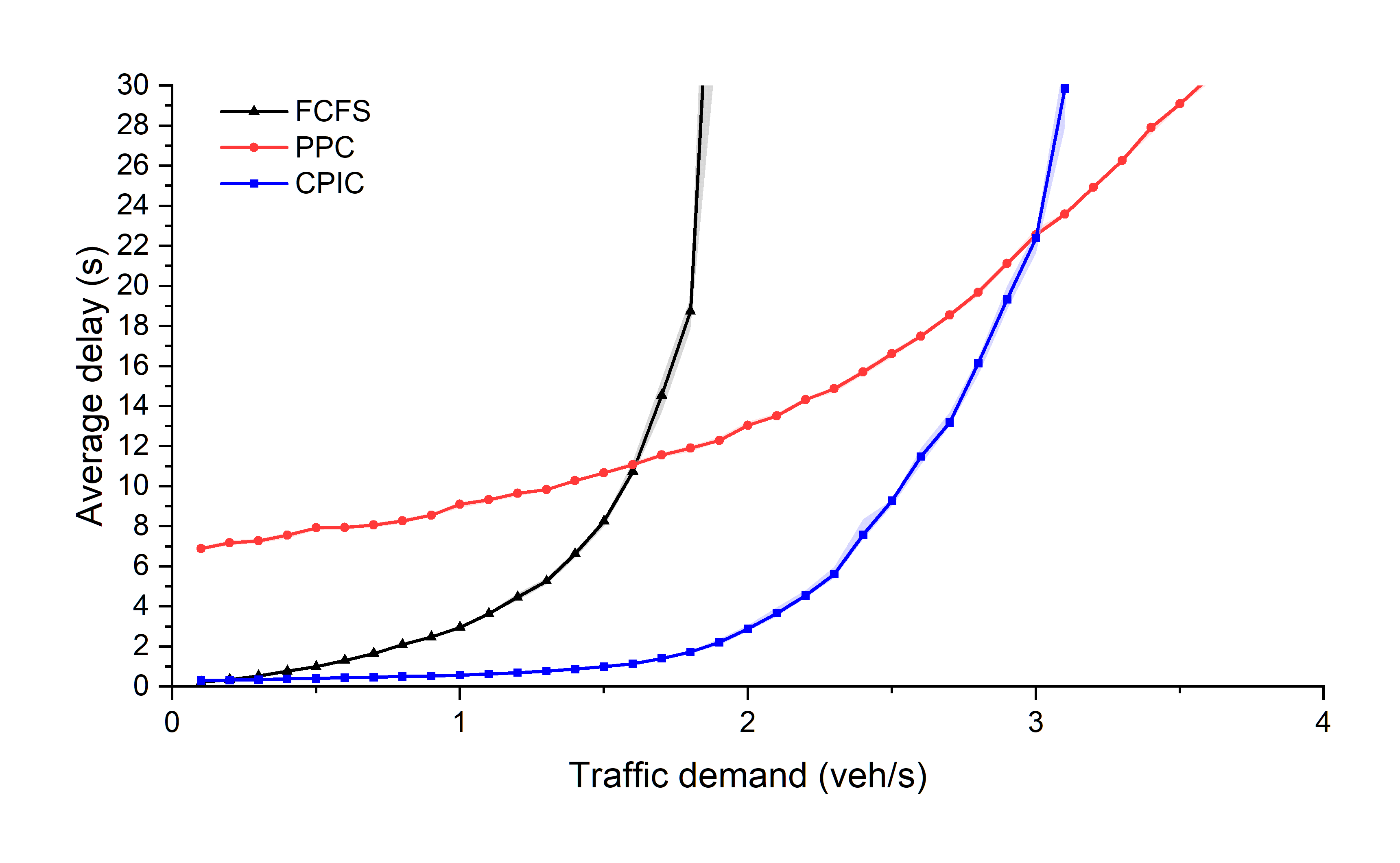}}
	\caption[]{Simulation results of scenarios \textbf{4-A}(a) and \textbf{4-B}(b)}
	\label{Fig15}
\end{figure}

From the simulation results presented in Figs.\ref{Fig15}(a) and \ref{Fig15}(b), it can be observed that the fluctuations in the traffic arrivals do not have much influence on the efficiency and capacity of these control models. As was revealed in scenarios \textbf{1-A} and \textbf{1-B}, the PPC strategy can handle higher traffic demand. On the other hand, FCFS and CPIC strategies are quite effective under low traffic demand, but their delay increases rapidly as the arrival rate $\lambda_0$ approaches their capacities.

\subsection{Intersection performance under different safety gaps} \label{simulation_Sce5_subsec}

Finally, we examine the intersection performance under different safety gap settings to reflect the impact of technological maturity of autonomous driving. In previous simulations, the minimum allowable gap between vehicles is 1.0 m. However, this setting requires a relatively high autonomous driving technology level, which is unlikely to be achieved in the near future. Therefore, to examine the impact of immature autonomous driving technologies, we conducted simulations with different safety gap settings under the same intersection layout and demand pattern of scenario \textbf{1-A}, and then compared the performance characteristics of the three control models. \par

\begin{figure}[!t]
    \centering
    \includegraphics[width=0.45\textwidth]{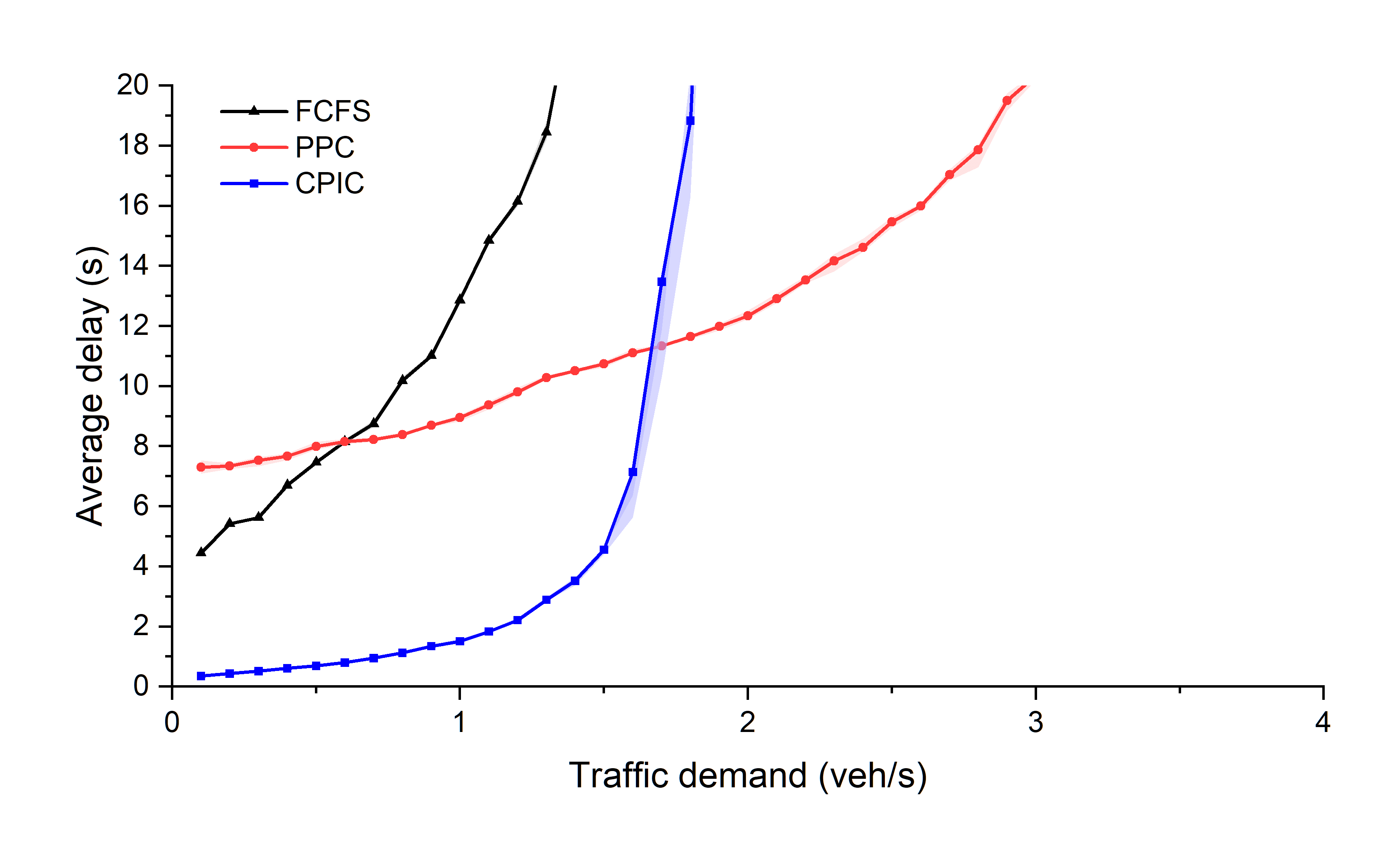}
    \caption{Simulation results under the safety gap of 4.0 meters}
    \label{Fig16}
\end{figure}

Fig. \ref{Fig16} illustrates the simulation results when the minimum allowable safety gap is set to 4.0 m. The change would have impacts on the headway of two consecutive vehicles from the same direction; furthermore, the minimum time gap between two conflict vehicles increases significantly, especially in conflict points with small angle $\theta$ (see Fig. \ref{Fig6}). As shown in Fig. \ref{Fig16}, the performance comparisons of the three control models follow similar patterns as in other scenarios: vehicle-based traffic control is dominant in the low demand cases, while the phase-based traffic control shows better performance in busy intersections. However, compared to the results of scenario \textbf{1-A} (shown in Fig. \ref{Fig9}(a)), the efficiency is reduced for all three control models. The delay under all the demand levels increases, and an excessive queuing time is observed under lower demand. For the FCFS strategy, the average delay begins to rise rapidly when $\lambda_0$ exceeds 1.4. The same result is obtained for the CPIC strategy (when $\lambda_0$ exceeds 1.8) and the PPC strategy (when $\lambda_0$ exceeds 3.5). Among the three control models, the PPC strategy is found to be affected the least by the increased safety gap. The critical traffic demand level $\lambda_0$ at which the PPC strategy outperforms the CPIC strategy is 1.7, which is much smaller than the value in scenario \textbf{1-A} (3.3). \par

\begin{figure}[!t]
    \centering
    \includegraphics[width=0.45\textwidth]{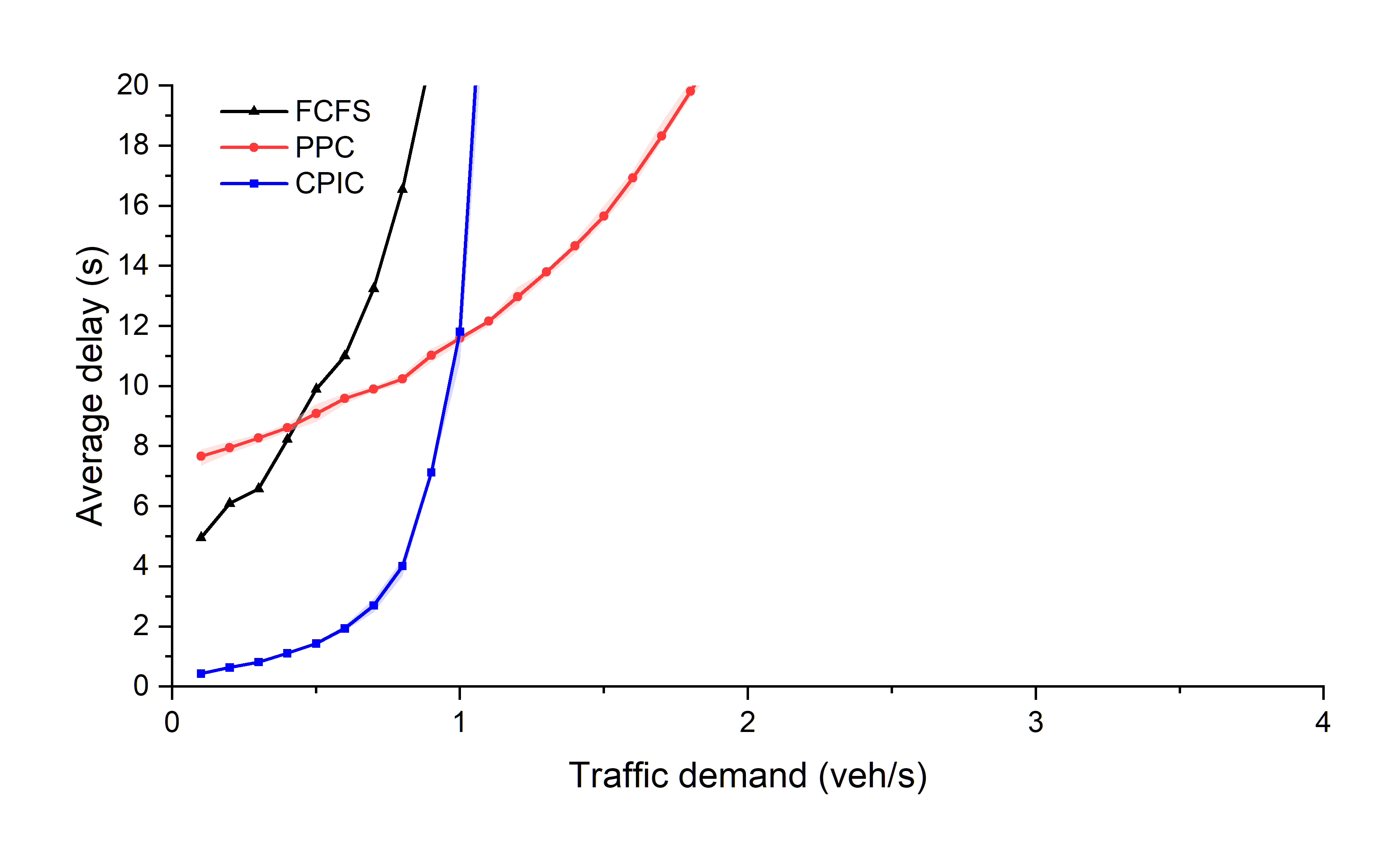}
    \caption{Simulation results under the safety gap of 8.0 meters}
    \label{Fig17}
\end{figure}

The results of the simulations conducted under a much higher safety gap setting are shown in Fig. \ref{Fig17}. When the minimum allowable safety gap is 8.0 m, the advantages of the PPC strategy are more distinct. The average delay of the vehicle-based traffic control exceeds the delay of the PPC strategy when $\lambda_0=1.0$, which is a quite low traffic demand level, indicating that the PPC may be much more suitable when autonomous driving technologies are not sufficiently advanced.

\section{Conclusions} \label{concluding_sec}

To supplement the existing studies on the isolated intersection control in the CAV era, this paper compared the performances of two intersection control philosophies, i.e., phase-based and vehicle-based traffic control, through a series of numerical simulations. Specifically, we implement three intersection control strategies: one ad-hoc negotiation based control~\cite{Dresner2008}, one rolling-horizon planning-based control (CPIC)~\cite{Levin2017Confilct} and one pretimed phase-based control. For a fair comparison of the three control strategies, all of the environmental factors, including the safe gap, the coordinating area length and the speed limits, were set to be the same to ensure that all three control strategies benefited equally from autonomous driving technologies. The comparisons were conducted under multiple intersection layouts, including symmetric and asymmetric 4-leg intersections as well as a T-type junction. In each layout, we simulated various traffic demand levels from 360 to 14,440 vehicles per hour and distributed the demand in both a balanced manner and an imbalanced manner. Furthermore, we compared the intersection performances under different settings, such as fluctuating vehicle arrival sequences and larger safe gaps. The simulation results lead to some interesting conclusions. Under the scenarios of 4-leg intersections, the vehicle-based traffic control strategies (FCFS and CPIC) show negligible delay when the traffic demand is low. As the demand level increases, their delay increases rapidly, making the phase-based traffic control the optimal approach. Nevertheless, in traffic scenarios with less conflicting vehicles (e.g., in a T-type junction with most vehicles driving on the main road), the vehicle-based methods show significantly improved performance. We also observe that when autonomous driving technologies are immature so that the CAVs are forced to maintain a larger headway, the advantages of phase-based traffic control over vehicle-based traffic control are much more distinct.  \par

This paper discussed several representative intersection control strategies in the autonomous driving era, and our comparisons cannot cover all possible control strategies; therefore, to acquire more reliable results, deriving the theoretical performances of intersection control under general settings is a topic worth investigation. On the other hand, the required safe gap is assumed to be fixed in this paper, which is not always reasonable, and the effects of a flexible gap should be examined in future studies. Finally, since some existing studies (e.g.,~\cite{Patel2019}) have revealed that the selection of control strategies at different intersections may affect the efficiency of traffic network, in the future it is necessary to generalize the comparisons to network level in order to obtain a comprehensive understanding of the performance of the network traffic control. \par

\section*{Acknowledgment}
The research is supported in part by Tsinghua-Daimler Joint Research Center for Sustainable Transportation and Tsinghua University-Toyota Research Center.

\ifCLASSOPTIONcaptionsoff
  \newpage
\fi

\begin{IEEEbiography}[{\includegraphics[width=1in,height=1.25in,clip,keepaspectratio]{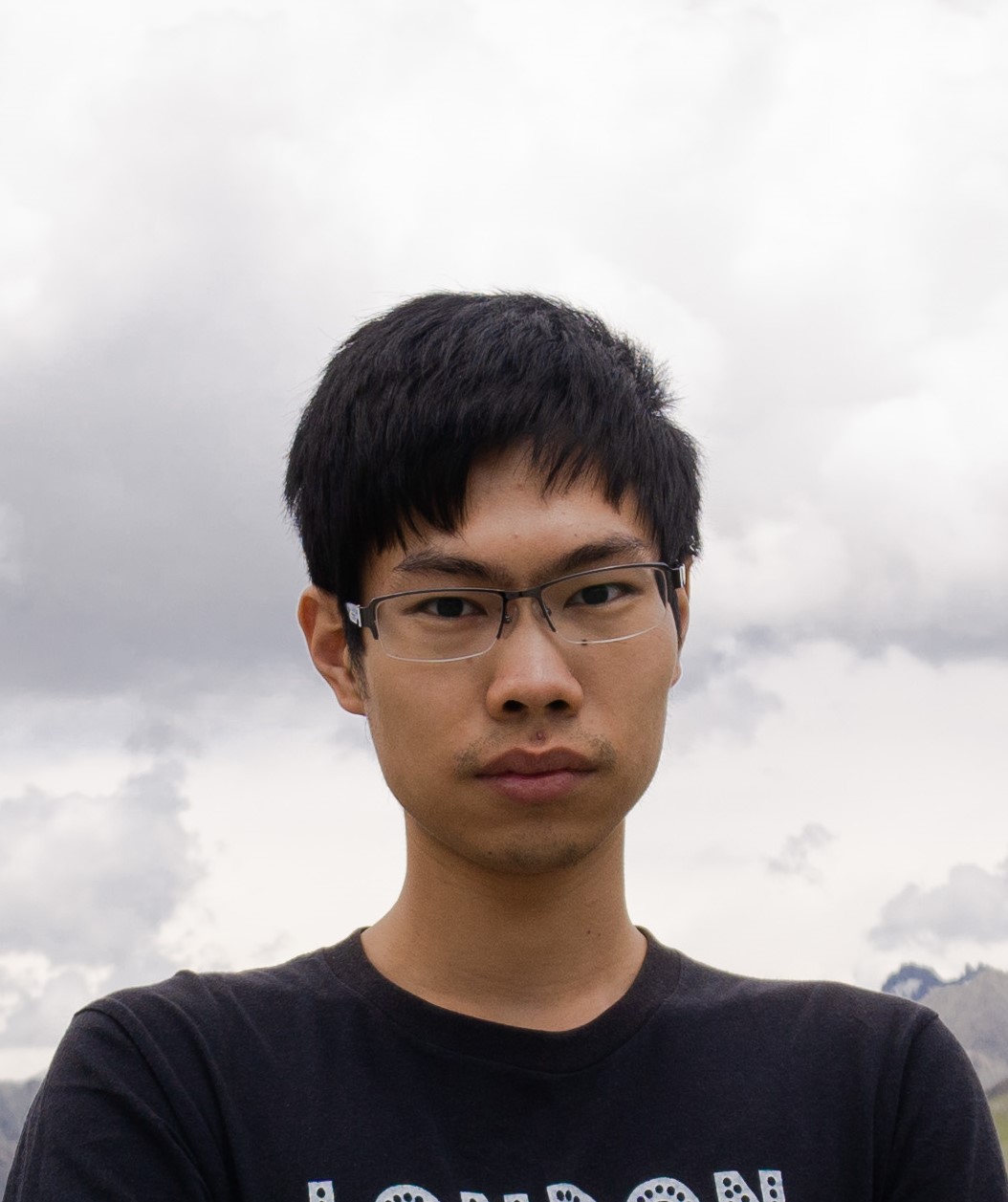}}]{Chen Yang}
is working toward the Ph.D. degree in the Department of Civil Engineering, Tsinghua University, Beijing, China, studying automated transportation systems.
\end{IEEEbiography}

\begin{IEEEbiography}[{\includegraphics[width=1in,height=1.25in,clip,keepaspectratio]{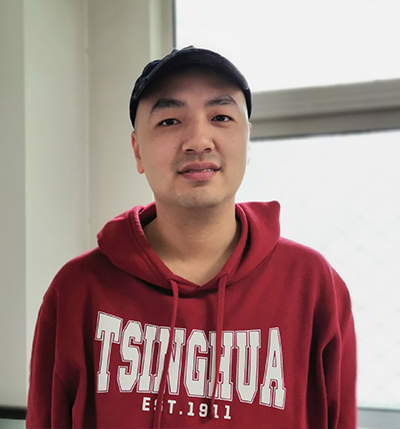}}]{Xi Lin}
is currently a research associate with the Department of Civil Engineering, Tsinghua University, Beijing, China, working in the fields of automated transportation systems, intelligent traffic control, electrified transportation and transportation economics. He had authored or coauthored more than 20 SCI indexed international journal papers and more than 15 international conference papers. Mr. Lin serves as the referee for multiple leading academic journals including IEEE Transactions on Intelligent Transportation Systems, Transportation Science, Transportation Research Part A/B/C/E and so on, and has won the acknowledgement of reviewing excellence on Transportation Research Part C.
\end{IEEEbiography}

%\newpage

\begin{IEEEbiography}[{\includegraphics[width=1in,height=1.25in,clip,keepaspectratio]{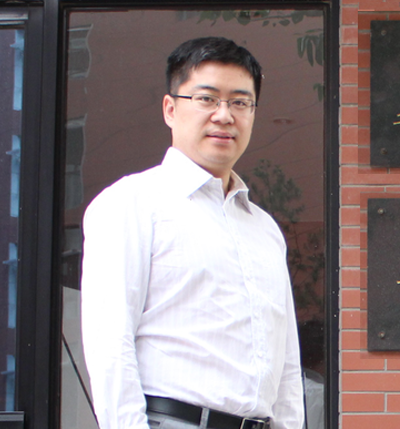}}]{Meng Li}
Meng Li received the B.S. degrees in structure engineering from Tsinghua University, China, in 2001, and the M.S. degrees in transportation science and supply chain management and M.E. degrees in transportation engineering from University of California, Berkeley, USA, in 2003 and 2004, respectively, and the Ph.D. degrees in transportation engineering from Nagoya University, Japan, in 2010.
He is an associate professor of Department of Civil Engineering with Tsinghua University, Beijing, China, working in the fields of transportation big data analysis, intelligent transportation systems and automated transportation.
Prof. Li serves as the editorial board of Journal of Intelligent Transportation Systems, the associate editor of Asia Pacific Journal of Operational Research, IEEE OpenJournal of Intelligent Transportation Systems and Smart and Resilient Transportation. He is also the general scientific committee member and topic area manager (TAM) of World Conference on Transport Research Society (WCTRS) and the committee member of sub-committee of Transportation Research Board (TRB), U.S.
\end{IEEEbiography}

\begin{IEEEbiography}[{\includegraphics[width=1in,height=1.25in,clip,keepaspectratio]{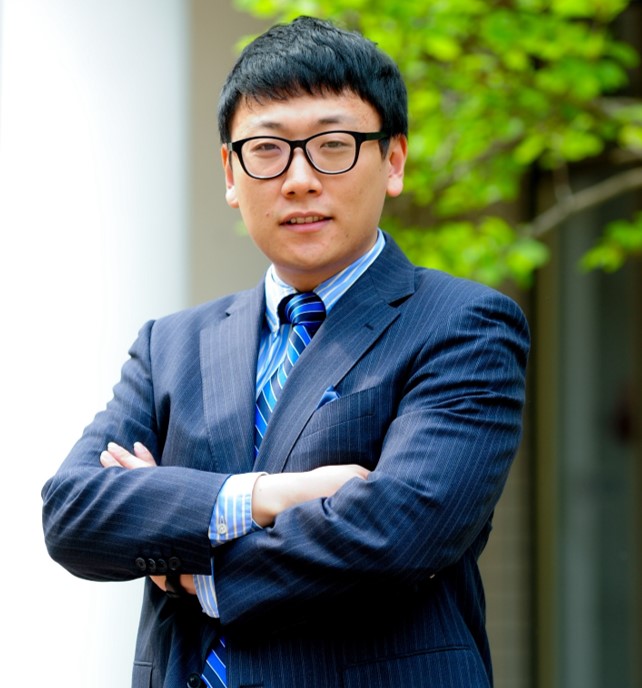}}]{Fang He}
Dr. Fang HE is currently an associate professor in the Department of Industrial Engineering, Tsinghua University, Beijing, China. His research focuses on comprehensively applying the knowledge of network modeling and optimization, machine learning, reinforcement learning, and economic analyses to solving management optimization problems associated with transportation science and transportation engineering. He currently serves as the editorial board member for Transportation Research Part C.
\end{IEEEbiography}


%\vfill


\end{document}